\documentclass{aa}  

\usepackage{natbib}
\bibpunct{(}{)}{;}{a}{}{,}

\usepackage{graphicx}
\usepackage{txfonts}
\newcommand{\xii}{\xi_{\mathrm{i}}}

\newcommand{\nuin}{\nu_{\mathrm{i H}}}
\newcommand{\nuni}{\nu_{\mathrm{H i}}}

\newcommand{\ca}{c_{\mathrm{A}}}

\newcommand{\nuine}{\nu_{\mathrm{i He}}}
\newcommand{\nunie}{\nu_{\mathrm{He i}}}

\newcommand{\nuh}{\nu_{\mathrm{H}}}
\newcommand{\nuhe}{\nu_{\mathrm{He}}}
\newcommand{\nui}{\nu_{\mathrm{i}}}

\newcommand{\rhoc}{\rho_{\rm c}}
\newcommand{\rhop}{\rho_{\rm p}}

\newcommand{\nunn}{\nu_{\mathrm{H He}}}
\newcommand{\nunne}{\nu_{\mathrm{He H}}}

\newcommand{\vi}{{\bf v}_{\rm i}}

\newcommand{\vn}{{\bf v}_{\rm H}}
\newcommand{\vne}{{\bf v}_{\rm He}}

\newcommand{\rhone}{\rho_{\rm He}}
\newcommand{\rhon}{\rho_{\rm H}}
\newcommand{\rhoi}{\rho_{\rm i}}

\newcommand{\ann}{\alpha_{\rm H He}}
\newcommand{\ain}{\alpha_{\rm i H}}
\newcommand{\aine}{\alpha_{\rm i He}}

\begin{document}

\title{The role of Alfv\'en wave heating in solar prominences}

\titlerunning{Alfv\'en wave heating in prominences}

   \author{Roberto Soler\inst{1,2}, Jaume Terradas\inst{1,2}, Ramon Oliver\inst{1,2}, and Jose~Luis Ballester\inst{1,2}}
   
   \authorrunning{Soler et al.}

   \institute{Departament de F\'isica, Universitat de les Illes Balears, E-07122 Palma de Mallorca, Spain.    
   \and
  Institut d'Aplicacions Computacionals de Codi Comunitari (IAC$^3$), Universitat de les Illes Balears, E-07122 Palma de Mallorca, Spain.  \\
\email{roberto.soler@uib.es} }

   \date{Received XXX; accepted XXX}

% \abstract{}{}{}{}{} 
% 5 {} token are mandatory
 
  \abstract
   {Observations have shown that magnetohydrodynamic waves over a large frequency range are ubiquitous in solar prominences. The waves are probably driven by photospheric motions and may transport energy up to prominences suspended in the corona. Dissipation of wave energy can lead to heating of the cool prominence plasma, so contributing to the local energy balance within the prominence. Here we discuss  the role of Alfv\'en wave dissipation as a   heating mechanism for the prominence plasma. We consider a slab-like quiescent prominence model with a transverse magnetic field embedded in the solar corona. The prominence medium is modelled as a  partially ionized  plasma composed of a charged ion-electron single fluid and two separate neutral fluids corresponding to neutral hydrogen and neutral helium. Friction between the three fluids acts as a dissipative mechanism for the waves.  The heating caused by  externally-driven Alfv\'en waves incident on the prominence slab is analytically explored. We find that the dense prominence slab acts as a resonant cavity for the waves. The fraction of incident wave energy that is channelled into the   slab strongly depends upon the wave period, $P$. Using typical prominence conditions, we obtain that wave energy trapping and associated heating are negligible when $P \gtrsim 100$~s, so that it is unlikely that those waves have a relevant influence on  prominence energetics. When $1$~s~$\lesssim P \lesssim 100$~s the energy absorption into the slab shows several sharp and narrow peaks, that can reach up to $\sim 100\%$, when the incident wave frequency matches a cavity resonance of the slab.  Wave heating is enhanced at those resonant frequencies. Conversely, when $P \lesssim 1$~s cavity resonances are absent, but the waves are heavily damped by the strong dissipation.  We estimate that  wave heating  may compensate for about 10\% of  radiative losses of the  prominence plasma.}


   \maketitle
%
%________________________________________________________________

\section{INTRODUCTION}

The energy balance in solar prominences, and the understanding of the processes involved with heating and cooling of the plasma, are difficult problems that are intimately linked to the prominence formation and structure \citep[see, e.g.,][]{gilbert2015,heinzel2015}. Using the differential emission measure, \citet{parenti2007} found that for a prominence the integral over the temperature range of $10^4$~K--$10^6$~K yielded radiative losses of $\sim 3\times 10^6$~ergs~cm$^{-2}$~s$^{-1}$. In quiescent conditions the prominence radiation is steady, requiring a source of still unknown heating to maintain energy balance in the structure \citep{parenti2014}. Identifying the source or sources of prominence heating is a challenging task for both observers and theoreticians, and several heating mechanisms have been proposed \citep[see][]{labrosse2010,Heinzel2010,gilbert2015}. 

Incident radiation is generally accepted as the dominant prominence heating mechanism \citep[see][]{gilbert2015}. The amount of radiative heating in prominences largely depends on the  illumination from the surrounding solar atmosphere. For this reason, although the prominence plasma is akin to that in the chromosphere, the studies of chromospheric radiative heating  \citep[see, e.g.,][]{carlsson2012} are not entirely applicable to the particular conditions of prominences.  Because prominences are suspended in and  surrounded by the corona, coronal radiation is absorbed from all sides, while they are somewhat more  isolated from the photospheric radiation than the chromospheric plasma \citep{gilbert2015}. The radiation field plays a crucial role and has to be properly considered when evaluating the balance between radiative heating and radiative losses \citep[see][]{heinzel2015}. 

Several works tried to construct prominence models by taking into account the balance between incident  radiation and cooling. \citet{Heasley1976} obtained radiative-equilibrium temperatures of about 4,600~K in slab-like prominences. \citet{Anzer1999} considered similar models and concluded that  some form of additional heating is necessary to balance radiative losses inside the prominence, specially in the prominence cool core. In addition, computations by \citet{Heinzel2010,Heinzel2012} point out that not only the radiation field is important but the accurate description of radiative losses is also crucial to estimate the radiative-equilibrium temperature in prominences. \citet{Heinzel2012} considered a more accurate description of prominence cooling by adding the calcium radiative losses and computed equilibrium temperatures  even lower than those obtained for a pure hydrogen gas. In general, the obtained radiative-equilibrium temperatures in the works cited above are lower than what is typically inferred from observations. The values reported in the literature, as derived applying different methods, span in a range of temperatures between 7,500~K and 9,000~K in the prominence cores \citep[see][]{parenti2014}. An additional, non-negligible source of heating seems to be necessary to rise the central prominence temperature up to the expected values \citep{labrosse2010,Heinzel2010,Heinzel2012}. In this regard, dissipation of magnetohydrodyamic (MHD) wave energy is proposed as another possible source of prominence heating to be considered in addition to the dominant radiative heating. The mechanism of wave heating in the solar atmosphere has been recently reviewed by \citet{arregui2015}. The actual importance of prominence wave heating, however, remains to be explored in detail.

High-resolution observations have unquestionably demonstrated the presence of  MHD waves in solar prominences \citep[e.g.,][]{lin2007,lin2009,hillier2013}. Also, non-thermal velocities derived from line widths  have been interpreted as a signature of unresolved waves \citep{parenti2007}. The non-thermal velocities observed by \citet{parenti2007} were about 10~km~s$^{-1}$ for prominence core temperatures. Statistical studies performed by \citet{hillier2013} on the properties of transverse waves in thin threads of a quiescent prominence revealed  wave periods ranging from 50~s to 6000~s, and  velocity amplitudes  between 0.2~km~s$^{-1}$ and 23~km~s$^{-1}$, with a significant amount of wave energy found at short periods. We note that, owing to observational constraints, periods shorter than 50~s could not be detected by \citet{hillier2013}, but it is likely that such short periods are also present in prominences \citep[see a discussion on this issue in][]{pecseli2000}. Indeed, periods of about 30~s have been reported in prominences from the analysis of Doppler time series \citep{Balthasar1993,Kolobov2008}. We also note that the typical lifetime of individual threads is $\sim$~20~min, so that the detection of longer periods in prominence fine structures should be interpreted with caution.   The driver of the  waves  has been related to photospheric granular motions, so that the driver would be an agent external to the prominence itself. In this direction, \citet{hillier2013} compared the power spectra of the prominence waves with that obtained from photospheric horizontal motions and obtained a good correlation. Thus, MHD waves might transport energy from the photosphere up to prominences suspended high in the corona. The question then arises, can this wave energy be deposited in the prominence plasma in the form of heat?

The observational evidence of MHD waves in prominences and their theoretical understanding are well established \citep[see][]{arreguireview,ballester2015}. However, detailed studies of the actual role of the waves in prominence heating are scarce and their conclusions are dissimilar. For instance, \citet{pecseli2000} studied the possible role of Alfv\'en wave dissipation in providing a supporting force for the prominence material against gravity \citep[see also][]{jensen1983,jensen1986} and estimated that the associated wave heating would only contribute with a minute fraction to the bulk radiative energy. Conversely, \citet{parenti2007}  assumed that the non-thermal velocities measured in a quiescent prominence were caused by the presence of Alfv\'en waves and performed a simple estimation of the Alfv\'en wave energy flux. Then, \citet{parenti2007} suggested that   radiative losses for prominence-corona transition region (PCTR) temperatures could be entirely compensated by Alfv\'en wave heating alone \citep[see also][]{parenti2014}. We must note, however, that \citet{parenti2007} probably overestimated the wave heating rate by assuming that all the wave energy flux is converted into heat. As pointed out by  \citet{gilbert2015},  detailed investigations of the impact of wave heating in prominences are necessary. 

The ability of MHD waves to heat the plasma relies on the presence of a dissipative process capable of efficiently damping the waves and depositing their energy into the medium. The efficiency of a given dissipative mechanism usually depends on the temporal and spatial scales involved \citep[see, e.g.,][]{soler2015apj}. A comparative study of the role of various damping mechanisms of MHD waves  was made by \citet{khodachenko2004,khodachenko2006}. In the case of plasmas with prominence conditions,  \citet{khodachenko2004,khodachenko2006} found that damping by ion-neutral collisions is dominant over other processes as, e.g., viscosity or magnetic resistivity. Thus, ion-neutral collisions can be the required mechanism for efficient dissipation of  MHD waves in prominences. 

The damping of MHD waves in partially ionized prominence plasmas has been theoretically investigated in a number of works \citep[e.g.,][among others]{forteza2007,soler2009PI,carbonell2010}, although no elaborated connection between wave damping and plasma heating has  been provided so far. In the solar chromosphere,  Alfv\'en wave damping due to ion-neutral collisions is a subject of intense research \citep[e.g.,][]{haerendel1992,depontieu2001,leake2005,soler2015}. It has been suggested that the heating associated to the Alfv\'en wave damping  may be strong enough to partly compensate chromospheric radiative losses \citep[see, e.g.,][]{goodman2011,song2011,russell2013,tu2013,reep2016,arber2016}. In view of the similarities between chromospheric and prominence plasmas, the purpose of this paper is to perform an exploratory study of the role of Alfv\'en waves in prominence heating. 

Another relevant question is whether the energy of  waves driven at the phostosphere can be efficiently channelled into prominences. Basic results on the reflection and transmission of Alfv\'en waves incident on a contact discontinuity \citep[see][]{ferraro1954} indicate that most of the  wave energy incident on prominences should be reflected because of the large prominence-to-corona density ratio. However, \citet{hollweg1984a} showed that the depression in the ambient Alfv\'en velocity caused by a dense plasma slab, such as a prominence, acts as a resonant cavity for  incident Alfv\'en waves \citep[see also][]{hollweg1981,ionson1982,Zhugzhda1982,hollweg1984b,Sterling1984}. \citet{hollweg1984a} applied his results to coronal loops, but the essential physics would be the same in the case of prominence slabs. Due to the existence of certain resonant frequencies associated to nearly standing waves in the cavity, large wave energy fluxes can be transmitted into the slab from the external medium, so enhancing the efficiency of wave energy transmission at those resonant frequencies.  The physical origin of the cavity resonances studied by \citet{hollweg1984a}  resides in the interference of the waves. Complex interference patterns can be formed because of the wave reflection and transmission. A resonance occurs when the waves transmitted into the cavity form a nearly standing wave and, in addition, there is a destructive interference between the reflected waves and the waves that leak out of the cavity. Thus, we note that  cavity resonances  have a different physical origin than the so-called Alfv\'en continuum resonances caused by the variation of the plasma properties across the magnetic field direction \citep[see, e.g.,][]{goossens2011}. Here we shall use a simple slab model with a transverse magnetic field to represent a  prominence embedded in the solar corona and shall explore the combined impact of cavity resonances and ion-neutral dissipation on the wave heating of prominences. To the best of our knowledge, the role of cavity resonances has not been investigated before in the case of prominences.

This paper is organized as follows. Section~\ref{sec:basic} contains the basic equations that govern the propagation of linear Alfv\'en waves in a partially ionized plasma. In Section~\ref{sec:slab} we embark on the  investigation of the heating by Alfv\'en waves in a prominence slab model. Finally, Section~\ref{sec:conclusion} contains the conclusions and some ideas for future works.

\section{BASIC EQUATIONS}
\label{sec:basic}

Here we give the general governing equations of linear Alfv\'en wave propagation in a homogeneous partially ionized prominence plasma with a straight and constant magnetic field, $\bf B$. We assume that the plasma constituents are hydrogen and helium. Because of the relatively low prominence core temperature ($T\lesssim 10^4$~K), we assume that hydrogen is only partially ionized while, for simplicity, helium is assumed to be fully neutral. The abundance of ionized helium at  prominence core temperatures is very small, so that its influence on the behavior of Alfv\'en waves is expected to be of minor importance.

\subsection{Multi-fluid plasma}

For the study of wave propagation, we consider  the so-called multi-fluid formalism \citep[see, e.g.,][]{zaqarashvili2011a}. It is assumed that  ions and electrons form a  ion-electron single fluid, while neutral hydrogen and neutral helium form two separate neutral fluids. We denote by $\rhoi$, $\rhon$, and $\rhone$ the density of ions, neutral hydrogen, and neutral helium, respectively.  Hence the total  density is
\begin{equation}
\rho \approx \rhoi + \rhon + \rhone,
\end{equation}
where the mass density of electrons is neglected owing to the very small electron mass. The hydrogen ionization ratio is
\begin{equation}
\xii = \frac{\rhoi}{\rhoi + \rhon}.
\end{equation}
We consider $\xii$ to be an arbitrary parameter whose value is  between $0 < \xii < 1$. The helium to hydrogen ratio of abundances by number is
\begin{equation}
\frac{A_{\rm He}}{A_{\rm H}} = \frac{n_{\rm He}}{n_{\rm i} + n_{\rm H}}, 
\end{equation}
where $n_{\rm i}$, $n_{\rm H}$, and $n_{\rm He}$ are the number densities of ions, neutral hydrogen, and neutral helium, respectively. We also consider $A_{\rm He}/A_{\rm H}$ to be an arbitrary parameter. Hence, the densities of the individual fluids as functions of the total density are
\begin{eqnarray}
\rhoi &=& \frac{\xii}{1+4A_{\rm He}/A_{\rm H}}\rho, \\
\rhon &=& \frac{1-\xii}{1+4A_{\rm He}/A_{\rm H}}\rho, \\
\rhone &=& \frac{4A_{\rm He}/A_{\rm H}}{1+4A_{\rm He}/A_{\rm H}}\rho.
\end{eqnarray}

 The various fluids  exchange momentum  by means of particle collisions.  Electron-neutral collisions are of minor importance for the present study  and are not taken into account here. The friction coefficient determines the strength of the  coupling between fluids. The expression of the friction coefficient for collisions involving neutrals in the approximation of small velocity drift is \citep[e.g.,][]{brag,draine86}
\begin{equation}
\alpha_{\rm \beta\beta'} = \frac{\rho_\beta\rho_{\beta'}}{m_{\beta}+m_{\beta'}} \frac{4}{3}\sigma_{\beta\beta'}\sqrt{\frac{8 k_{\rm B} T}{\pi m_{\beta\beta'}}}
\end{equation}
where $\beta$ and $\beta'$ stand for i, H, or He, $T$ is the temperature (we assume the same temperature for all the components of the plasma), $m_\beta$ is the particle mass of species $\beta$, $m_{\beta\beta'} = m_{\beta} m_{\beta'}/ \left( m_{\beta} + m_{\beta'} \right)$ is the reduced mass, $k_{\rm B}$ is Boltzmann's constant, and $\sigma_{\beta\beta'}$ is the momentum-transfer cross section.  We take $\sigma_{\rm iH} \approx 10^{-18}$~m$^{-2}$ and $\sigma_{\rm iHe} \approx 3\times10^{-19}$~m$^{-2}$ from \citet{vranjes2013}, and $\sigma_{\rm HHe} \approx 1.5\times 10^{-19}$~m$^{-2}$ from \citet{Lewkow2012}.

\subsection{Governing equations of linear  waves}

To study linear waves, we assume small perturbations superimposed on the background plasma and linearize the general, non-linear multi-fluid equations \citep[see, e.g.,][]{zaqarashvili2011a,Zaqarashvili2011b,Khomenko2014}.  For simplicity, we ignore the effect of gravity and also assume that there are no equilibrium flows. The linearized equations considered here for the study of  Alfv\'en waves are
\begin{eqnarray}
\rhoi \frac{\partial \vi}{\partial t} &=& \frac{1}{\mu} \left( \nabla \times {\bf b} \right) \times {\bf B} - \ain \left( \vi - \vn \right) \nonumber \\
&&- \aine \left( \vi - \vne \right), \label{eq:momlinion} \\
\rhon \frac{\partial \vn}{\partial t} &=& - \ain \left( \vn - \vi \right) - \ann \left( \vn - \vne \right), \label{eq:momlinh} \\
\rhone \frac{\partial \vne}{\partial t} &=& - \aine \left( \vne - \vi \right) - \ann \left( \vne - \vn \right), \label{eq:momlinhe}\\
\frac{\partial {\bf b}}{\partial t} &=& \nabla \times \left( \vi \times {\bf B} 	\right), \label{eq:induction}  
\end{eqnarray}
where $\vi$, $\vn$ and $\vne$ are the velocity perturbations of ions, neutral hydrogen, and neutral helium, respectively,  ${\bf b}$ is the magnetic field perturbation,  $\mu$ is the magnetic permeability, and the rest of parameters have been defined before. In these equations we have not considered gas pressure perturbations because they are irrelevant for Alfv\'en waves. Other effects neglected in these equations are viscosity and magnetic resistivity, whose role on the wave damping has been shown to be less important than that of ion-neutral collisions in a partially ionized prominence plasma \citep{khodachenko2004,khodachenko2006}. The induction equation (Equation~(\ref{eq:induction})) is derived from the generalized Ohm's Law obtained after neglecting electron inertia. The full expression of the generalized Ohm's Law in a partially ionized plasma can be found in, e.g.,  \citet{zaqarashvili2011a,Khomenko2014,soler2015}. Here, we use the ideal version of Ohm's Law, in which the term corresponding to electron drift (i.e., Hall's effect) is neglected. This term has been shown to be unimportant concerning the behavior of  MHD waves in a partially ionized prominence plasma \citep[see][]{soler2009PIRA}. Specifically, Hall's  effect is  negligible for wave frequencies lower than the ion cyclotron frequency, namely $\Omega_{\rm i} = e B/m_{\rm i}$, with $B$  the magnetic field strength. For a typical prominence magnetic field strength of 10~G, we have $\Omega_{\rm i} \approx 96$~kHz, which is much higher than the frequency of the observed waves in prominences \citep[see, e.g.,][]{hillier2013}.

\subsection{Wave energy  and  heating rate}

An expression for the evolution of the energy of the linear waves can be obtained by following a similar procedure to that described in, e.g., \citet{goedbloed2004} and    \citet{walker2005} for the case of ideal MHD waves in a fully ionized plasma. We take the dot product of Equations~(\ref{eq:momlinion}), (\ref{eq:momlinh}), and (\ref{eq:momlinhe}) with $\vi$, $\vn$, and $\vne$, respectively, and also take the dot product of Equation~(\ref{eq:induction}) with ${\bf b}/\mu$. Next, we add the equations and the resulting expression governing the wave energy evolution can be cast as
\begin{equation}
\frac{\partial U}{\partial t} + \nabla \cdot {\bf \Pi} = - Q, \label{eq:energycon}
\end{equation}
where $U$, ${\bf \Pi}$, and $Q$ are given by
\begin{eqnarray}
U  &=& \frac{1}{2} \rhoi \left| \vi \right|^2 + \frac{1}{2} \rhon  \left| \vn \right|^2 + \frac{1}{2} \rhone  \left| \vne \right|^2  + \frac{1}{2\mu}\left| {\bf b} \right|^2 , \label{eq:energydensity} \\
{\bf \Pi}  &=& \frac{1}{\mu} \left[  \left({\bf B} \cdot {\bf b} \right) \vi - \left(\vi \cdot {\bf b} \right){\bf B} \right], \label{eq:energyflux} \\
Q &=& \ain \left| \vi - \vn \right|^2 +  \aine \left| \vi - \vne \right|^2 +  \ann \left| \vn - \vne \right|^2. \label{eq:heat}
\end{eqnarray}
Equation~(\ref{eq:energycon}) shows that the wave energy is not conserved. 

We discuss the meaning of the various terms in Equation~(\ref{eq:energycon}). The first term on the left-hand side accounts for the temporal evolution of the  wave energy density, $U$, which is the sum of the kinetic energies of ions, neutral hydrogen, and neutral helium, plus the magnetic energy. The second term  on the left-hand side is the divergence of the wave energy flux, $\bf \Pi$, that informs us about the amount of energy that propagates with the wave.  We note that only the velocity of  ions  appears in the expression of $ {\bf \Pi}$, so that Equation~(\ref{eq:energyflux}) is formally identical to that in fully ionized  plasmas \citep[see][]{goedbloed2004}. The right-hand side of Equation~(\ref{eq:energycon}) depends on the quantity $Q$ (Equation~(\ref{eq:heat})), which accounts for the loss of wave energy owing to dissipation due to collisions \citep[see also][]{brag,draine86}.

It is implicit in  Equation~(\ref{eq:energycon}) that, as the  wave propagates, its energy is absorbed into the plasma, where it is eventually transformed into heat.  The energy lost by the wave must be gained by the plasma and converted into internal energy. This could be seen in the full version of the  energy equation, in which  the changes of the background plasma were also included \citep[see][for the energy equation in a fully ionized  plasma]{goedbloed2004}. The full energy equation remains a conservation equation. Wave energy is converted into internal energy of the background plasma, but the sum is constant. The plasma heating itself cannot be studied with the present linear analysis, but Equation~(\ref{eq:energycon}) informs about the rate at which the wave energy is lost. We can therefore identify $Q$ as the wave heating rate.

\subsection{Alfv\'en wave dispersion relation}

We consider monochromatic, parallel propagating  Alfv\'en waves and set  the perturbations as proportional to $\exp\left(-i\omega t + i k s \right)$, where $\omega$ is the wave frequency, $k$ is the wavenumber in the direction of the background magnetic field, and $s$ denotes the coordinate along the background magnetic field lines. The frequency, $\omega$, is assumed to be real. Because of wave  damping, the wavenumber, $k$, is complex. The imaginary part of the wavenumber defines the characteristic damping length of the waves.  

For incompressible Alfv\'en waves, Equations~(\ref{eq:momlinion})--(\ref{eq:induction}) can be recast as follows,
\begin{eqnarray}
\omega  \vi &=&   - \frac{k B}{\mu\rhoi} {\bf b} - i \nuin \left( \vi - \vn \right)  - i \nuine \left( \vi - \vne \right), \label{eq:momional}\\
\omega \vn &=&  - i \nuni \left( \vn - \vi \right) - i \nunn \left( \vn - \vne \right), \label{eq:momneual} \\
\omega  \vne &=&  - i \nunie \left( \vne - \vi \right)  - i \nunne \left( \vne - \vn \right), \label{eq:momneuale} \\
\omega {\bf b} &=&   - k B\, \vi , \label{eq:inductal}
\end{eqnarray}
where  $\nuin$, $\nuni$, $\nuine$, $\nunie$, $\nunn$, and $\nunne$ are  collision frequencies  defined as
\begin{eqnarray}
\nuin &=& \frac{\ain}{\rhoi}, \qquad \nuni = \frac{\ain}{\rhon}, \\
\nuine &=& \frac{\aine}{\rhoi}, \qquad \nunie = \frac{\aine}{\rhone} \\
\nunn &=& \frac{\ann}{\rhon}, \qquad \nunne = \frac{\ann}{\rhone}.
\end{eqnarray}
From Equations~(\ref{eq:momneual}) and (\ref{eq:momneuale}) we find that the velocities of neutral hydrogen and neutral helium are related to the ion velocity by
\begin{eqnarray}
 \vn &=&   \frac{i \nuni \left( \omega + i\nuhe \right) - \nunn \nunie}{\left( \omega + i\nuh \right)\left( \omega + i\nuhe \right) + \nunn\nunne}    \vi, \label{eq:alfv2} \\
 \vne &=&  \frac{i \nunie \left( \omega + i\nuh \right) - \nunne \nuni}{\left( \omega + i\nuh \right)\left( \omega + i\nuhe \right) + \nunn\nunne}  \vi, \label{eq:alfv3}
\end{eqnarray}
with $\nuh = \nuni + \nunn$ and $\nuhe = \nunie + \nunne$ the  neutral hydrogen and neutral helium total collision frequencies, respectively.

We combine Equation~(\ref{eq:momional})  with Equations~(\ref{eq:inductal}), (\ref{eq:alfv2}), and (\ref{eq:alfv3}) to remove the other perturbations and arrive at an equation for $\vi$ alone (not given here). The condition that $\vi \ne \bf 0$ provides us with the dispersion relation, namely
\begin{equation}
k^2 = \frac{\omega \Omega_{\rm col}}{\ca^2}, \label{eq:reldispera}
\end{equation}
where $\ca^2$ is the square of the ion Alfv\'en velocity (which only includes the ion density) given by
\begin{equation}
\ca^2 = \frac{B^2}{\mu \rhoi},
\end{equation}
and $\Omega_{\rm col}$ is a redefined, complex-valued frequency that incorporates the effects of collisions with neutrals and is given by
\begin{eqnarray}
\Omega_{\rm col} & \equiv & \omega + i \nui +  \frac{\nuin \nuni \left( \omega + i\nuhe \right) + i \nuin \nunn \nunie}{\left( \omega + i\nuh \right)\left( \omega + i\nuhe \right) + \nunn\nunne} \nonumber \\
&&+   \frac{\nuine \nunie \left( \omega + i\nuh \right) + i \nuine \nunne \nuni}{\left( \omega + i\nuh \right)\left( \omega + i\nuhe \right) + \nunn\nunne}, \label{eq:omegacol}
\end{eqnarray}
where $\nui=\nuin + \nuine$ is the  ion total collision frequency.  In the case that the plasma is fully ionized, $\Omega_{\rm col} = \omega $ and Equation~(\ref{eq:reldispera}) simplifies to the well-known dispersion relation of  ideal Alfv\'en waves, namely
\begin{equation}
k^2 = \frac{\omega^2}{\ca^2}. \label{eq:reldisperideal}
\end{equation}
The wavenumber is real in a fully ionized ideal plasma.

\subsection{Wave energy deposition}
\label{sec:damping}

We use Equation~(\ref{eq:energyflux}) to compute the Alfv\'en wave energy flux. Because of the Fourier dependence $\exp\left(-i\omega t + i k s \right)$ the wave variables are complex and the physical meaningful part resides in the real part. In addition,  to drop the oscillatory behavior of the waves and  retain the net energy flux, we average $\bf \Pi$ over one wave period, $P=2\pi/\omega$. Thus, the time-averaged energy flux, $\left< {\bf \Pi} \right> $, for a forward-propagating wave is
\begin{equation}
\left< {\bf \Pi} \right> =  \frac{B^2}{2\mu} \frac{ {\rm Re}\left( k \right)}{\omega} V^2 \exp\left[-2 {\rm Im} \left(k\right) s \right] \hat{e}_B, \label{eq:alfenergyflux}
\end{equation}
where  $\hat{e}_B$ is the unit vector in the direction of the background magnetic field, $\rm Re$ and $\rm Im$  denote the real and imaginary parts, respectively,  $k$ is given by the principal square root of Equation~(\ref{eq:reldispera}), and $V$ is the  amplitude of the ion velocity perturbation at the arbitrary position $s=0$. Both ${\rm Re}\left( k \right)$ and ${\rm Im}\left( k \right)$ are positive. The units of $\left< {\bf \Pi} \right>$ are W~m$^{-2}$. Equation~(\ref{eq:alfenergyflux}) agrees with Equation~(8) of \citet{hollweg1984a}.

To check Equation~(\ref{eq:alfenergyflux}) we consider  a fully ionized plasma, so that ${\rm Im} \left(k\right)=0$ and ${\rm Re} \left(k\right)=\omega/\ca$. Then, Equation~(\ref{eq:alfenergyflux}) consistently reverts to the well-known expression of the ideal Alfv\'en wave energy flux \citep[e.g.,][]{walker2005}, namely
\begin{equation}
\left< {\bf \Pi} \right> =  \frac{1}{2} \rhoi V^2 \ca  \hat{e}_B. 
\end{equation}

 Equation~(\ref{eq:alfenergyflux}) evidences that the net energy transported by the Alfv\'en wave exponentially decreases as the wave propagates through the plasma. This behavior is a consequence of the damping due to  collisions. A characteristic energy absorption exponential length scale, $L_A$, can be defined as
\begin{equation}
L_A = \frac{1}{2{\rm Im} \left(k\right)}. \label{eq:la}
\end{equation} 
The wave energy is absorbed, i.e., deposited into the plasma in a length scale of the order of $L_A$.

\begin{figure}[!tb]
\centering
\includegraphics[width=.95\columnwidth]{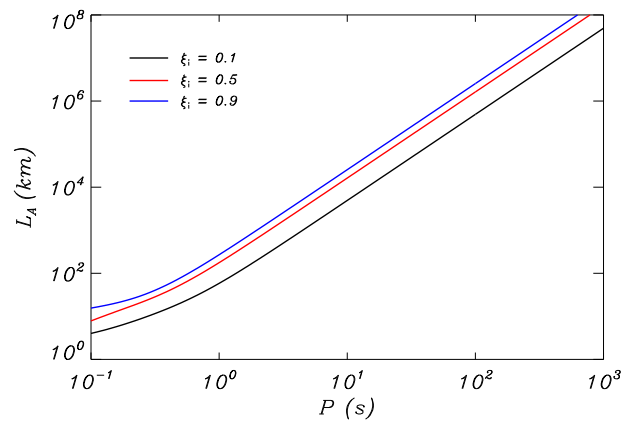}
\caption{Energy absorption length, $L_A$, as a function of the wave period, $P=2\pi/\omega$, for an Alfv\'en wave propagating in a  prominence plasma when different values of the hydrogen ionization ratio, $\xii$, are assumed. The meaning of the various linestyles is indicated within the figure.  The parameters used in the computations are $\rho = 5\times 10^{-11}$~kg~m$^{-3}$, $T=8000$~K, $B=10$~G, and  $A_{\rm He}/A_{\rm H}=0.1$. We note that logarithmic scale has been used in both axes. \label{fig:homogeneous}}
\end{figure}

In order to estimate $L_A$, we consider typical physical conditions in quiescent  prominences. We take $\rho = 5\times 10^{-11}$~kg~m$^{-3}$, $T=8000$~K, and $B=10$~G. In addition, we take  $A_{\rm He}/A_{\rm H}=0.1$ \citep{Gilbert2002}. Figure~\ref{fig:homogeneous} displays the energy absorption length, $L_A$, computed from Equation~(\ref{eq:la}) as a function of the wave period, $P$, for various values of the hydrogen ionization ratio, $\xii$. As expected, the lower the  hydrogen ionization ratio, the stronger the wave damping because of the larger presence of neutrals in the plasma. Waves with short periods are  very efficiently damped in the prominence plasma. The corresponding energy absorption lengths for these short-period waves are rather small. For instance, $L_A$ is between 100~km and 1000~km for waves with periods between 1~s and 10~s, and $L_A$ keeps decreasing when $P$ is further reduced. Conversely, the wave damping, and so the energy absorption, is less efficient for waves  with longer periods. We obtain that $L_A$ increases as the wave period is increased. For instance, for wave periods of the order of 100~s, the energy absorption length is as large as $10^6$~km. In practice, this result suggests that the energy deposition associated to  waves with periods of the order of minutes is not significant over realistic length scales  in solar prominences \citep[see, e.g,][]{engvold2015}. We shall confirm this result later.

To determine the relative importance of neutral helium on the wave damping, we have repeated the computation of Figure~\ref{fig:homogeneous} but when collisions with neutral helium are neglected. For the sake of simplicity, these results are not shown here. We obtain that collisions with helium are of minor importance in determining the value of  $L_A$ because collisions of ions with neutral hydrogen is  the dominant damping mechanism of the waves.

\section{WAVE HEATING IN A PROMINENCE SLAB}
\label{sec:slab}

In Section~\ref{sec:damping} we have learnt that short-period Alfv\'en waves can efficiently damp in prominence plasmas. Here, we investigate whether the energy of Alfv\'en waves incident on a solar prominence can be deposited in the prominence medium because of this damping. To do so, we represent the prominence by a slab with a transverse magnetic field embedded in the solar corona and assume the presence of Alfv\'en waves that are incident on the prominence-corona interface (see Figure~\ref{fig:model}). In this work we do not care about the driver of the waves, which has been associated to  photospheric motions \citep{hillier2013}. We assume, however, that the wave driver is external to the prominence. Here we are interested in studying the interaction of the incident waves with the dense and cool, partially ionized slab.  

The incident waves will be partly transmitted into the  prominence slab and partly reflected back to the corona. The waves that penetrate into the prominence will travel back and forth through the whole prominence slab, and part of their energy will be leaked again to the  corona. The superposition of these transmitted and reflected waves can cause complex interference patterns. Previous investigations by \citet{hollweg1984a} showed that the dense slab acts as a resonant cavity for the incident Alfv\'en waves. Cavity resonances may appear when the waves transmitted into the slab interfere so that a nearly standing wave is formed and, at the same time, the waves reflected back to the corona and the waves that leak out of the slab interfere destructively. In such a case, most of the incident wave energy is channelled into the slab. In addition, since the waves are  dissipated in the partially ionized prominence plasma because of ion-neutral collisions, the wave energy fed into the cavity resonances may efficiently contribute to the heating of the plasma.

Essentially, the condition for the excitation of cavity resonances is that the frequency of the incident waves matches a natural frequency (eigenmode) of the prominence. The eigenmodes of a prominence modelled as a dense plasma slab have been investigated in the literature \citep[e.g.,][]{joarder1992,oliver1993}, where the dependence of the natural frequencies on the physical and geometrical properties of the slab was explored. Here, we shall give an approximate expression for the Alfv\'en cavity frequencies in Section~\ref{sec:cavity}. The excitation of an internal Alfv\'en eigenmode naturally produces large wave amplitudes within the prominence slab. In the stationary state, the energy of the wave driver must be distributed among the internal and external plasmas. Hence, if a cavity resonance is hit, most of the available energy goes to the internal plasma and, consequently, the amount of energy in the exterior is much smaller. Physically, the low wave amplitude in the external medium can be understood in terms of the destructive interference mentioned before.

\subsection{Slab model}
\label{sec:model}

\begin{figure}[!t]
\centering
\includegraphics[width=.95\columnwidth]{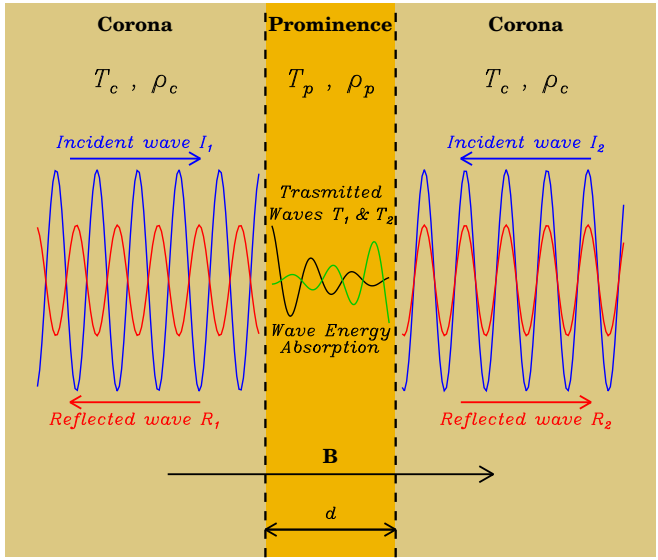}
\caption{Sketch of the prominence slab model. \label{fig:model}}
\end{figure}

We assume that the equilibrium state is composed of a uniform partially ionized plasma slab of width $d$, representing a solar prominence, embedded into a uniform and unbounded fully ionized medium representing the solar coronal environment. When necessary, we shall use the subscripts `p' and `c' to  refer to the plasma in the prominence slab and in the corona, respectively. We use Cartesian coordinates and set the coordinate frame so that the equilibrium is infinite in the $y$- and $z$-directions, while the prominence slab is bounded in the $x$-direction. The center of the slab is located at $x=0$. The equilibrium  is permeated by a straight magnetic field along the $x$-direction, namely ${\bf B} =  B \hat{e}_x$, so that the magnetic field is transverse to the prominence slab. Similar  slab models have been used in the past to investigate global oscillations (eigenmodes) of prominences \citep[see, e.g.,][]{joarder1992,oliver1993,anzer2009,soler2009NewA}.  A sketch of the slab model can be seen in Figure~\ref{fig:model}.

Since the equilibrium magnetic field is uniform,  the  thermal pressure is required to be continuous across the prominence-corona interface. We use the ideal gas law to compute the  coronal, $P_{\rm c}$, and prominence, $P_{\rm p}$, equilibrium pressures as
\begin{equation}
P_{\rm c} = \frac{\rhoc R T_{\rm c}}{\tilde{\mu}_{\rm c}}, \qquad P_{\rm p} = \frac{\rhop R T_{\rm p}}{\tilde{\mu}_{\rm p}},
\end{equation}
while the requirement that the  thermal pressure is continuous across the prominence-corona interface gives the relation between the densities and temperatures of the two plasmas as
\begin{equation}
\frac{\rhoc}{\rhop} = \frac{T_{\rm p}/\tilde{\mu}_{\rm p}}{T_{\rm c}/\tilde{\mu}_{\rm c}}. \label{eq:pressbalance}
\end{equation}
In these equations, $R$ is the ideal gas constant,  and $\tilde{\mu}_{\rm c}$ and $\tilde{\mu}_{\rm p}$ are the mean atomic weight in the corona and in the prominence, respectively. On the one hand, the coronal plasma is fully ionized so that
\begin{equation}
\tilde{\mu}_{\rm c}= \frac{1+4\frac{A_{\rm He}}{A_{\rm H}}}{2+3\frac{A_{\rm He}}{A_{\rm H}}}. \label{eq:mutcorona}
\end{equation}
Assuming a typical helium abundance of $A_{\rm He}/A_{\rm H} = 0.1$, Equation~(\ref{eq:mutcorona}) results in $\tilde{\mu}_{\rm c} \approx 0.6$. On the other hand, in the prominence we consider the hydrogen ionization degree  to be arbitrary and  helium to be fully neutral. Hence,
\begin{equation}
\tilde{\mu}_{\rm p}= \frac{1+4\frac{A_{\rm He}}{A_{\rm H}}}{1+\xii + \frac{A_{\rm He}}{A_{\rm H}}}.  \label{eq:mutprom}
\end{equation}
For instance, if $A_{\rm He}/A_{\rm H} = 0.1$ and $\xii = 0.5$, Equation~(\ref{eq:mutprom}) gives $\tilde{\mu}_{\rm p} = 0.875$.

\subsection{Matching conditions at the prominence-corona boundary}

We consider  Alfv\'en waves propagating in the corona that are incident on the prominence-corona interface. Since both the prominence slab and the corona are homogeneous, the Alfv\'en wave dispersion relations (Equation~(\ref{eq:reldispera}) in the prominence and Equation~(\ref{eq:reldisperideal}) in the corona) are locally satisfied.  The incident waves are partly transmitted into the  prominence slab and partly reflected back to the corona. Here we find the velocity amplitudes of the transmitted and reflected waves in terms of the amplitudes of the incident waves. The process is similar to that followed by \citet{hollweg1984a}.

In the steady state, and assuming that the waves are plane polarized along the $y$-direction, we can write the ion velocity perturbation as
\begin{equation}
v_{{\rm i}y} \left( x \right) = \left\{ 
\begin{array}{lll}
I_1 \exp\left(i k_{\rm c} x \right) + R_1 \exp\left(-i k_{\rm c} x \right), & \textrm{if} & x < -\frac{d}{2}, \\
T_1  \exp\left(i k_{\rm p} x\right) + T_2 \exp\left(- i k_{\rm p} x\right), & \textrm{if} &  |x| < \frac{d}{2}, \\
I_2 \exp\left( -i k_{\rm c} x \right) + R_2 \exp\left( i k_{\rm c} x \right), & \textrm{if} & x > \frac{d}{2},
\end{array}
  \right.
\end{equation}
where $I_1$ and $I_2$ are the amplitudes of the incident waves, $R_1$ and $R_2$ are the amplitudes of the waves reflected back to the corona, and $T_1$ and $T_2$ are the amplitudes of the waves transmitted into the prominence slab. In addition, $k_{\rm c}$ and $k_{\rm p}$ are the parallel wavenumber in the corona and in the prominence, respectively. On the one hand, $k_{\rm c}$ is given by taking the positive square root of the right-hand side of Equation~(\ref{eq:reldisperideal}) and using coronal conditions. On the other hand, $k_{\rm p}$ is given by taking the principal square root of the right-hand side of Equation~(\ref{eq:reldispera}) and using prominence conditions. 

We assume that the amplitudes of the incident waves, $I_1$ and $I_2$, are arbitrary. Then, the amplitudes $R_1$, $R_2$, $T_1$, and $T_2$ are determined by imposing appropriate boundary conditions at $x=\pm d/2$. In the case of an Alfv\'en wave incident on a contact discontinuity, the boundary conditions are the continuity of the tangential components of the magnetic and  electric fields \citep[see, e.g.,][]{goedbloed2004}. In our case, these conditions simply reduce to the requirements
\begin{equation}
\left[\left[ v_{{\rm i}y}\right]\right] = 0, \qquad \left[\left[ \frac{\partial v_{{\rm i}y}}{\partial x} \right]\right] = 0, \label{eq:boundary}
\end{equation}
where $\left[\left[ X \right]\right]$ denotes the variation of the quantity $X$ across the prominence corona-interface.

We impose the boundary conditions given in Equation~(\ref{eq:boundary}) at $x=\pm d/2$ and, after some algebraic manipulations, we find that the amplitudes of the reflected and transmitted waves as functions of $I_1$ and $I_2$ are
\begin{eqnarray}
R_1 &=&  \frac{\frac{4 k_{\rm p} k_{\rm c}}{\left( k_{\rm p} + k_{\rm c} \right)^2} I_2 +  \frac{2 i \left(k_{\rm p}^2 - k_{\rm c}^2\right)}{\left( k_{\rm p} + k_{\rm c} \right)^2} \sin\left(k_{\rm p} d \right) I_1}{1-\left( \frac{k_{\rm p} - k_{\rm c}}{k_{\rm p} + k_{\rm c}} \right)^2\exp\left( i 2 k_{\rm p} d \right)}\exp\left[ i \left( k_{\rm p} - k_{\rm c} \right) d\right], \\
R_2 &=&  \frac{\frac{4 k_{\rm p} k_{\rm c}}{\left( k_{\rm p} + k_{\rm c} \right)^2} I_1 +  \frac{2 i \left(k_{\rm p}^2 - k_{\rm c}^2\right)}{\left( k_{\rm p} + k_{\rm c} \right)^2} \sin\left(k_{\rm p} d \right) I_2}{1-\left( \frac{k_{\rm p} - k_{\rm c}}{k_{\rm p} + k_{\rm c}} \right)^2\exp\left( i 2 k_{\rm p} d \right)}\exp\left[ i \left( k_{\rm p} - k_{\rm c} \right) d\right], \\
T_1 &=&  \frac{\frac{2  k_{\rm c}}{k_{\rm p} + k_{\rm c}} I_1 +  \frac{2 k_{\rm c} \left(k_{\rm p} - k_{\rm c}\right)}{\left( k_{\rm p} + k_{\rm c} \right)^2} \exp\left(i k_{\rm p} d \right) I_2}{1-\left( \frac{k_{\rm p} - k_{\rm c}}{k_{\rm p} + k_{\rm c}} \right)^2\exp\left( i 2 k_{\rm p} d \right)}\exp\left[ i \left( k_{\rm p} - k_{\rm c} \right) \frac{d}{2}\right], \label{eq:t1} \\
T_2 &=&  \frac{\frac{2  k_{\rm c}}{k_{\rm p} + k_{\rm c}} I_2 +  \frac{2 k_{\rm c} \left(k_{\rm p} - k_{\rm c}\right)}{\left( k_{\rm p} + k_{\rm c} \right)^2} \exp\left(i k_{\rm p} d \right) I_1}{1-\left( \frac{k_{\rm p} - k_{\rm c}}{k_{\rm p} + k_{\rm c}} \right)^2\exp\left( i 2 k_{\rm p} d \right)}\exp\left[ i \left( k_{\rm p} - k_{\rm c} \right) \frac{d}{2}\right]. \label{eq:t2}
\end{eqnarray}

We note that \citet{hollweg1984a} only considered wave incidence on the left interface of the slab, so his case can be recovered by setting $I_2 = 0$ in our expressions. The amplitudes of the reflected and transmitted waves are complex valued and depend upon the wave frequency through the expression of the wavenumbers. It is worth noting that in the absence of dissipation the amplitudes of the reflected and transmitted Alfv\'en waves on a contact discontinuity are real and independent of the frequency \citep[see, e.g.,][]{ferraro1954}.  In general, the consideration of dissipative mechanisms  causes the  amplitudes of the reflected and transmitted waves to depend upon the wave frequency.

\subsection{Energy absorption coefficient}
\label{sec:abs}

The fraction of incident wave energy that is reflected  back to the corona is determined by the coefficient of energy reflectivity defined as
\begin{equation}
\mathcal{R} = \frac{\left| \left< {\bf \Pi} \right> \right|_{R_1} + \left| \left< {\bf \Pi} \right> \right|_{R_2}}{\left| \left< {\bf \Pi} \right> \right|_{I_1} + \left| \left< {\bf \Pi} \right> \right|_{I_2}}, 
\end{equation}
where $\left| \left< {\bf \Pi} \right> \right|_{I_1}$ and $\left| \left< {\bf \Pi} \right> \right|_{I_2}$ denote the moduli of the time-averaged energy flux of the incident waves, and $\left| \left< {\bf \Pi} \right> \right|_{R_1}$ and $\left| \left< {\bf \Pi} \right> \right|_{R_2}$ are the corresponding quantities of the reflected waves. The energy flux is given by  Equation~(\ref{eq:alfenergyflux}). In terms of the wave amplitudes, the expression of the reflectivity  simply becomes
\begin{equation}
\mathcal{R}  = \frac{R_1 R_1^* + R_2 R_2^*}{I_1 I_1^* + I_2 I_2^*},
\end{equation}
where the superscript $*$ denotes the complex conjugate. We assume that, in general, the amplitudes of the incident waves, $I_1$ and $I_2$, are complex valued, so that we can consider an arbitrary phase relation between the two incident waves. The values of $\mathcal{R} $ range between $\mathcal{R} =0$ (no reflection) and $\mathcal{R} =1$ (full reflection). In addition, the fraction of incident wave energy that is transmitted and trapped into the prominence slab can be computed by invoking conservation of energy.  So, the fraction of energy absorbed into the prominence is
\begin{equation}
\mathcal{A} = 1 -\mathcal{R}.
\end{equation}
The values of $\mathcal{A} $ range  between $\mathcal{A} =0$ (no energy absorption) and $\mathcal{A} =1$ (complete absorption). 

To avoid confusion, we must warn the reader that the present definitions of the reflection and absorption coefficients are different from those used by \citet{hollweg1984a}. We define the reflectivity as the fraction of incident energy that returns to the corona, and the absorption as the fraction of incident energy that gets trapped within the slab. \citet{hollweg1984a} only considered incidence on one boundary of the slab and defined the transmissivity as the fraction of incident wave energy that emerges from the opposite slab boundary.

\begin{figure*}[!t]
\centering
\includegraphics[width=1.95\columnwidth]{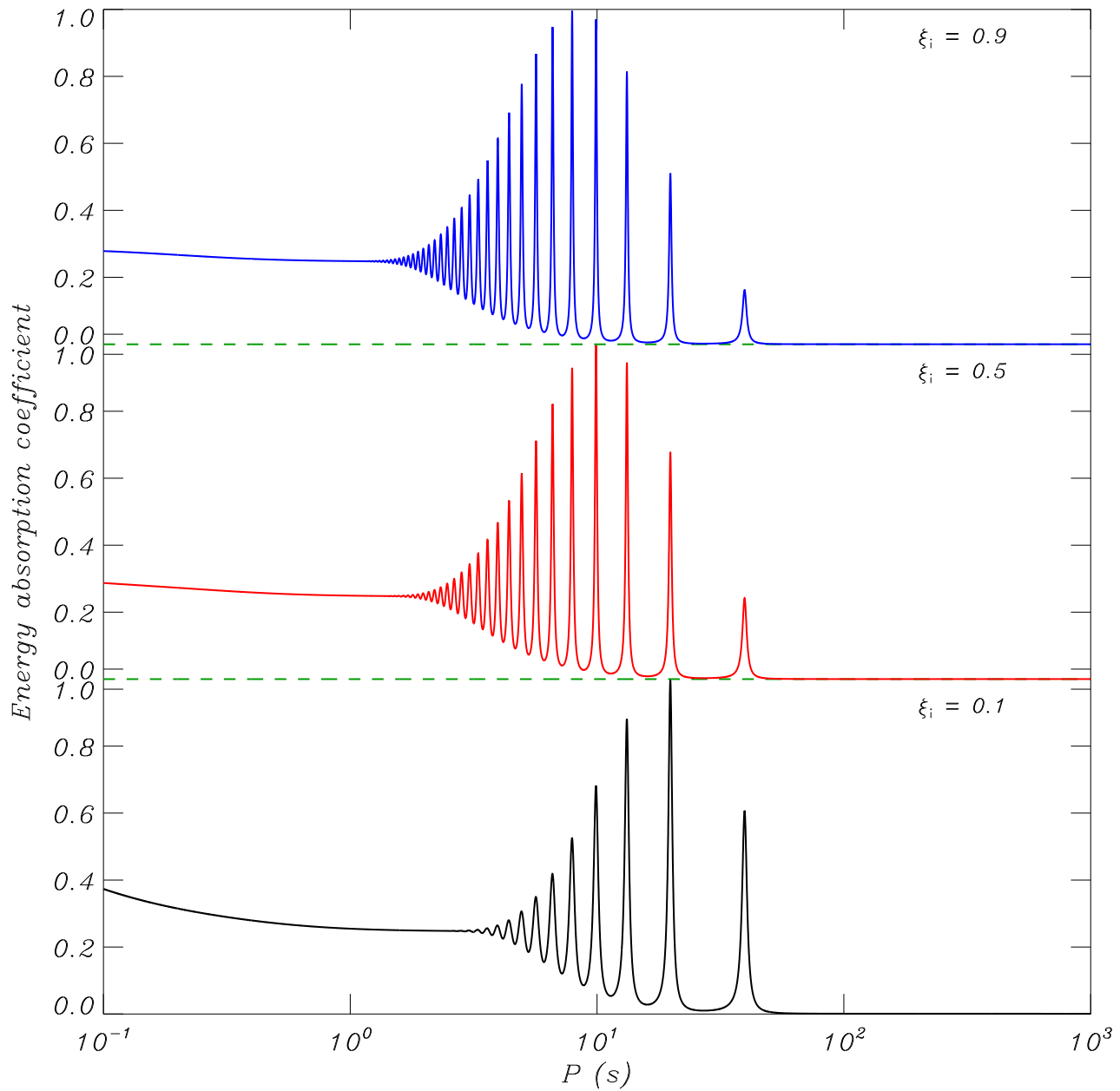}
\caption{Coefficient of energy absorption, $\mathcal{A}$, as a function of the wave period, $P=2\pi/\omega$, for various prominence hydrogen ionization ratio: $\xii=0.9$ (top graphs), $\xii=0.5$ (mid graphs), and $\xii=0.1$ (bottom graphs). We considered $I_1 = I_2$ and $d=$~5,000~km. We note that the horizontal axis is in logarithmic scale. The parameters used in the computations are given in the text. \label{fig:slab}}
\end{figure*}

\begin{figure*}[!t]
\centering
\includegraphics[width=1.95\columnwidth]{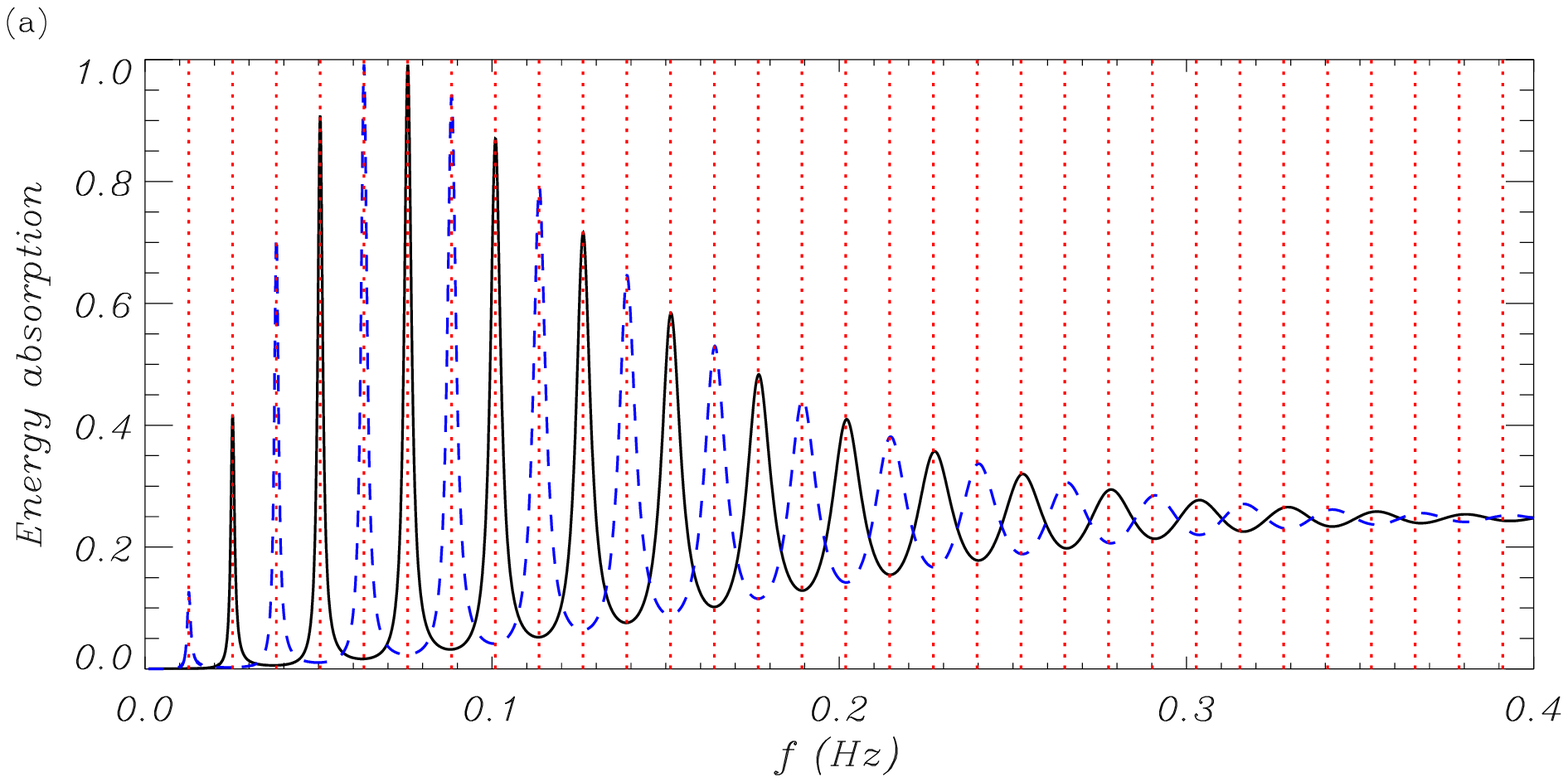}
\includegraphics[width=1.95\columnwidth]{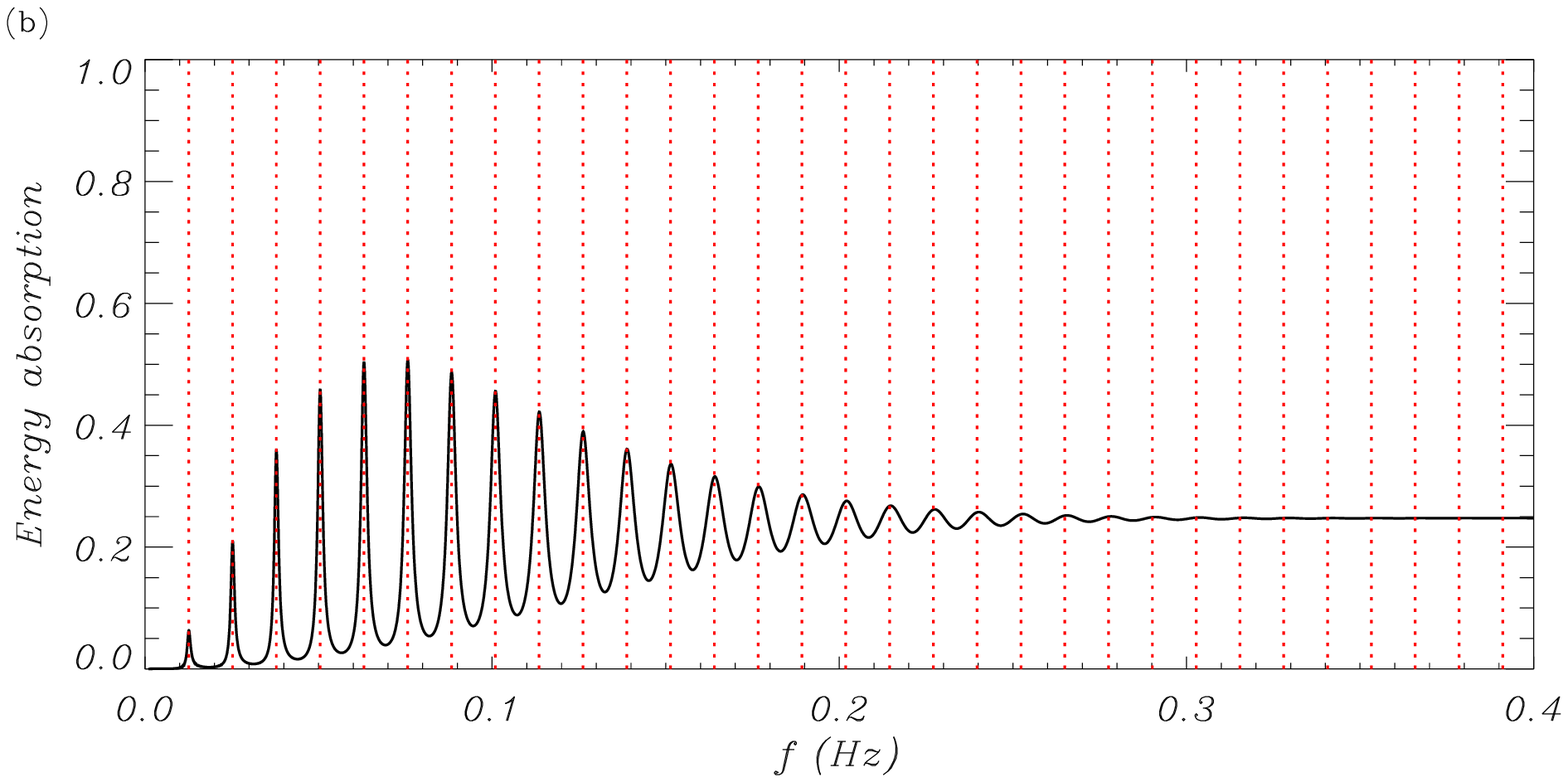}
\caption{ Coefficient of energy absorption, $\mathcal{A}$, as a function of the incident wave frequency (in Hz), $f=\omega/2\pi$. (a) Results when the incident waves on both sides of the slab are in phase ($I_1=I_2$, black solid line) and anti-phase ($I_1=-I_2$, dashed blue line). (b) Results when  $I_2 = I_1 e^{i \pi/2}$. In both panels, the vertical dotted lines denote the approximate cavity resonance frequencies according to Equation~(\ref{eq:natfreq}). In all cases we used $\xii=0.2$ and the remaining parameters are the same as in Figure~\ref{fig:slab}. \label{fig:slabdet}}
\end{figure*}

To perform a numerical application, we take $B=10$~G and  $A_{\rm He}/A_{\rm H}=0.1$. In the prominence plasma, we use the same physical conditions as before, namely $\rho_{\rm p} = 5\times 10^{-11}$~kg~m$^{-3}$ and $T_{\rm p}=8000$~K. We assume the density in the corona to be $\rho_{\rm c} = \rho_{\rm p}/200 = 2.5\times 10^{-13}$~kg~m$^{-3}$, and the coronal temperature is computed from the pressure balance condition (Equation~(\ref{eq:pressbalance})). For instance, if $\xii=0.5$, we get $T_{\rm c} \approx 1.1$~MK. We note, however, that the coronal temperature plays no role in the following results. In addition,  we take $d=$~5,000~km as the width of the prominence slab.

Figure~\ref{fig:slab} displays the coefficient of energy absorption, $\mathcal{A}$, for a wide range of incident wave periods (in logarithmic scale). The results for three different values of $\xii$ have been stacked one on top of the other for better comparison. In these computations we considered $I_1 = I_2$. The logarithmic scale used in Figure~\ref{fig:slab} evidences the existence of three different regimes depending upon the wave period:
\begin{itemize}
\item The large-period range, in which the energy trapping into the prominence slab is negligible and practically all the wave energy returns to the corona. This regime is independent of the value of $\xii$ and happens roughly for $P\gtrsim 100$~s.
\item The intermediate-period range, roughly for $1$~s~$\lesssim P \lesssim 100$~s, which is characterized by the presence of sharp absorption peaks that can reach up to $\mathcal{A} \sim 1 $. The location of the absorption peaks is independent of $\xii$. However, the height, width, and number of  peaks seem to be affected by $\xii$. We can relate the presence of these absorption peaks with the cavity resonances studied by \citet{hollweg1984a}. We shall explore the physics behind cavity resonances in Section~\ref{sec:cavity}.
\item The short-period range, approximately for $P \lesssim$~1~s, where the absorption stabilizes at about $\mathcal{A} \approx 0.3 $ and slowly grows when $P$ decreases. No absorption peaks are present here and the value of $\xii$ does not have an important effect.
\end{itemize}

The results of Figure~\ref{fig:slab} indicate that it is unlikely that waves with $P\gtrsim 100$~s impact on prominence energetics. The energy of these large-period waves is not trapped within the prominence. On the other hand, although the energy trapping of short-period waves with $P \lesssim$~1~s is more important, most of the energy is reflected back to the corona. Conversely, the existence of large  absorption peaks in the intermediate range of periods indicates that those resonances may efficiently trap energy within the slab. Cavity resonances are, therefore, excellent candidates to be involved in strong wave heating.

Importantly, we note that the coefficient of energy absorption as a function of the frequency is an intrinsic property of the model for a given set of parameters. The energy absorption does not depend on the velocity amplitude of the incident waves.

\subsection{Cavity resonances}
\label{sec:cavity}

 \citet{hollweg1984a} provided an instructive study about the nature of cavity resonances. We can follow  \citet{hollweg1984a} to understand the physics behind cavity resonances. According to \citet{hollweg1984a}, a resonance occurs when the waves transmitted into the slab interfere so that a nearly standing wave is formed. We can obtain an approximation to the resonant-cavity frequencies by neglecting the effect of dissipation (i.e., assuming a perfect coupling between the fluids). The approximation  can be cast as
\begin{equation}
\omega_n \approx \frac{n \pi}{d} \frac{B}{\sqrt{\mu\rho_{\rm p}}}, \qquad \textrm{with} \qquad n=1,2,3\dots \label{eq:natfreq}
\end{equation}
To derive this approximation we have assumed that the standing Alfv\'en wave is perfectly trapped within the slab.

Equation~(\ref{eq:natfreq}) corresponds to the frequencies of the internal Alfv\'en eigenmodes of the slab. The eigenmodes of a prominence slab embedded in the solar corona have been investigated by, e.g., \citet{joarder1992} and \citet{oliver1993}. In addition to the internal eigenmodes, there are external and hybrid modes that depend on the line-tying conditions at the base of the corona.   Since in the present model the corona is assumed to be unbounded, external and hybrid modes are absent. Anyway, neither hybrid nor external modes are expected to play an important role in prominence heating. Hybrid modes have smaller frequencies (longer periods) than all internal modes, so that their damping by ion-neutral collisions would be inefficient. In turn, the amplitude of external modes and their associated energy within the prominence slab is very small. 

Figure~\ref{fig:slabdet}(a) shows the coefficient of energy absorption as a function of the incident wave frequency (in Hz), $f=\omega/2\pi$, for a particular set of parameters given in the caption of the figure. In that figure, we use the frequency and not the period in the horizontal axis to better distinguish the resonance peaks. We use vertical lines to indicate the location of the first 31 resonances approximately given by Equation~(\ref{eq:natfreq}). We see that the location of the absorption peaks agrees very well with the approximate formula.  In Figure~\ref{fig:slabdet}(a) we consider two different cases, namely $I_1=I_2$ and $I_1 = - I_2$. The first case assumes that incident waves on both sides are in phase, while in the second case the incident waves are in anti-phase. These two cases are considered to illustrate that the symmetry of the incident waves determines whether their interference within the slab is constructive or destructive, so that only the resonances with the appropriate symmetry are excited. We see in Figure~\ref{fig:slabdet}(a) that only odd resonances (i.e., $n=1,3,5\dots$) are excited when $I_1 = - I_2$, while only even resonances (i.e., $n=2,4,6\dots$) appear when $I_1=I_2$. 

The two cases considered in Figure~\ref{fig:slabdet}(a) represent paradigmatic situations. In reality, we should not expect any special relation between the amplitudes and phases of the  waves incident on both sides of the slab. Therefore, odd and even resonances would be naturally excited by an arbitrary driver. This is seen in Figure~\ref{fig:slabdet}(b), where we took $I_2 = I_1 e^{i \pi/2}$. We note that the height of the absorption peaks in Figure~\ref{fig:slabdet}(b) is essentially half that of the peaks in Figure~\ref{fig:slabdet}(a). This is so because  the total number of  resonances is twice that of the paradigmatic cases of Figure~\ref{fig:slabdet}(a). Hence, wave energy absorption is now distributed among twice as many absorption peaks.

\begin{figure*}[!htp]
\centering
\includegraphics[width=1.95\columnwidth]{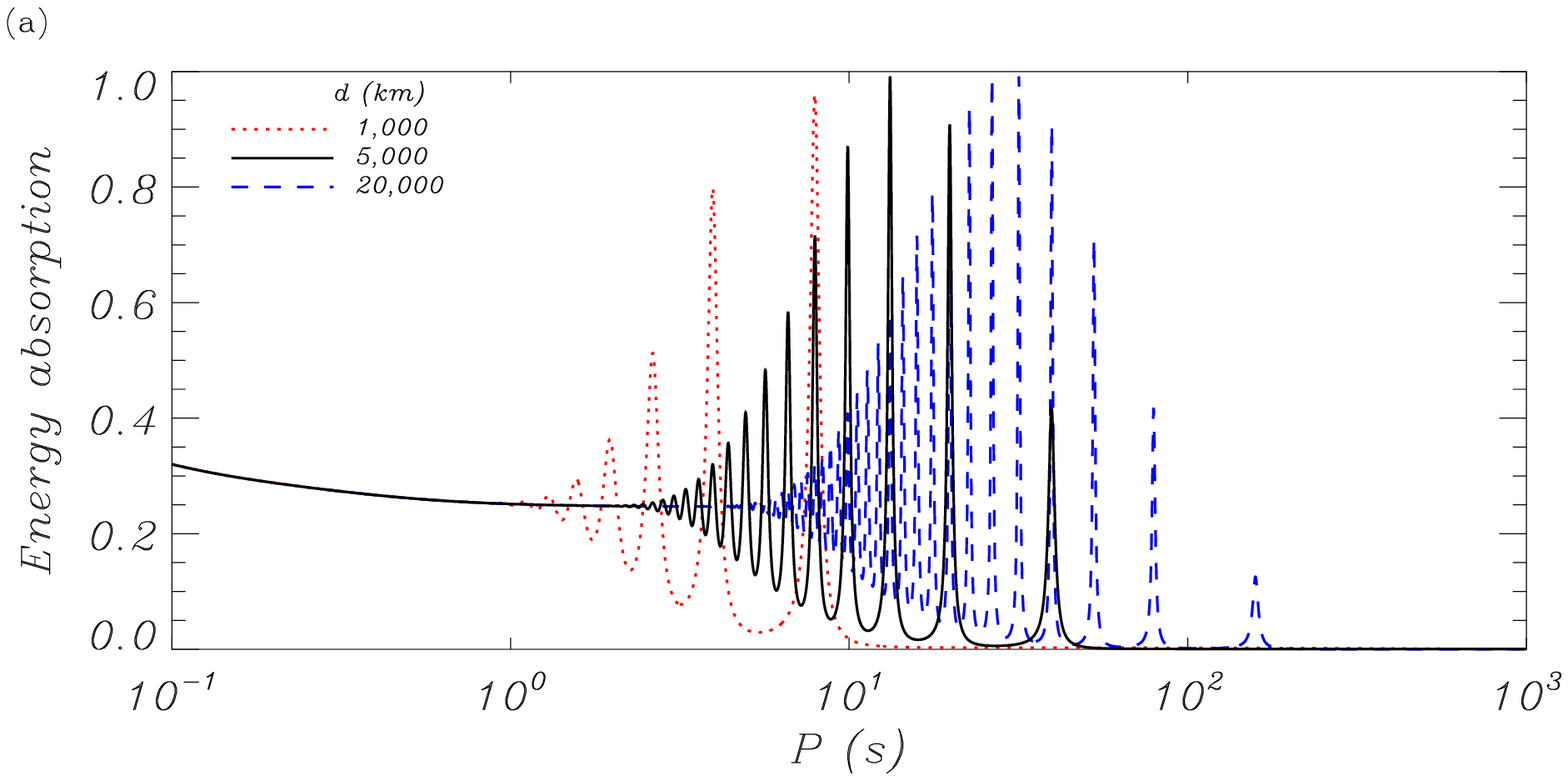}
\includegraphics[width=1.95\columnwidth]{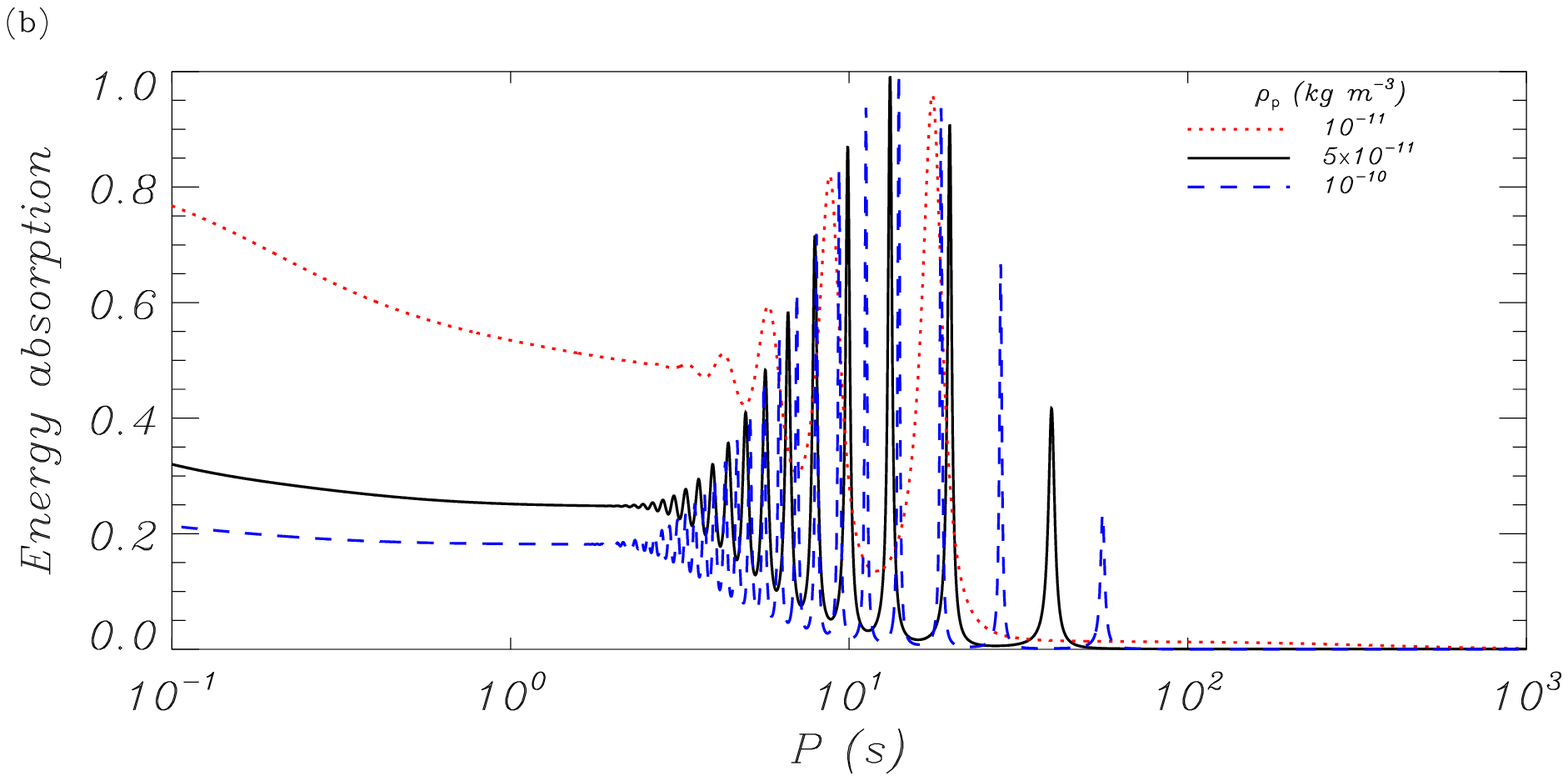}
\caption{Coefficient of energy absorption, $\mathcal{A}$, as a function of the incident wave period for (a) various values of the slab thickness, $d$, and (b) various values of the prominence density, $\rho_{\rm p}$. In all cases we used $\xii=0.2$, while the remaining parameters are the same as in Figure~\ref{fig:slab}. \label{fig:slab2}}
\end{figure*}

Equation~(\ref{eq:natfreq}) also informs us about the dependence of the resonant frequencies on the prominence density, $\rho_{\rm p}$, the magnetic field strength, $B$, and the width of the prominence slab, $d$. Figure~\ref{fig:slab2} shows the coefficient of energy absorption for various values of the slab width (panel a) and the prominence density (panel b). Both results are in agreement with Equation~(\ref{eq:natfreq}).  The decrease/increase of $d$ causes the resonant peaks to move to shorter/longer periods. Changing the density has a similar impact, but the density also influences the overall reflectivity of the slab, as happens in the case of a single discontinuity \citep{ferraro1954}. The smaller the  prominence density, the smaller the reflectivity, and so the larger the fraction of energy that gets trapped within the slab. This effect is specially noticeable in the short-period regime, but the density does not influence much the quality of the resonances.

\citet{hollweg1984a} showed that, in the absence of dissipation, leakage of the nearly standing wave to the external medium due to imperfect trapping determines the width and the maxima of the resonance peaks. The more important the leakage, the broader the resonance width and the smaller the resonance peak. Then, \citet{hollweg1984a} further added the effect of an arbitrary dissipation mechanism and obtained that dissipation has a similar effect as that of leakage on the shape of the resonances.

The results of \citet{hollweg1984a} can be used to qualitatively explain the shape of the absorption resonances obtained in the present work. Again, we use Figure~\ref{fig:slabdet} to support the discussion. Dissipation by ion-neutral collisions increases as the wave frequency increases. Accordingly, \citet{hollweg1984a} suggests that the width of the resonances should grow as the frequency of the resonance increases. Figure~\ref{fig:slabdet} clearly demonstrates this effect: the resonance peaks get  wider with increasing frequency. Concerning the peaks height, we see that the first six resonance peaks (i.e., those with the lowest frequencies) grow when the frequency increase. Conversely, the height of the resonance peaks with $n>6$ decrease as the frequency increases. According to \citet{hollweg1984a}, dissipation would be responsible for the decrease of the resonance peaks as the frequency increases. In turn, the increase of the height of the low-frequency resonances could be attributed to leakage becoming less efficient as the frequency increases.

A side effect of the decrease of the resonance height when the frequency increases is that resonances are in practice suppressed for high enough frequencies or, equivalently, short enough periods. This explains why resonance peaks are absent in the short-period range of Figure~\ref{fig:slab}.  The very strong dissipation that occurs for high frequencies (short periods) has the negative consequence of suppressing the cavity resonances and so reducing the efficiency of wave energy absorption into the slab. As $\xii$ increases, dissipation becomes more important and the effective number of resonances decreases.

\subsection{Associated heating}

The presence of cavity resonances in the intermediate range of periods dramatically enhances the transmission and trapping of wave energy into the prominence slab compared to the classic result of wave incidence on a simple plasma discontinuity \citep{ferraro1954}. The purpose here is to determine whether the ion-neutral dissipation of the  wave energy can provide significant plasma heating.

\begin{figure*}[!htp]
\centering
\includegraphics[width=1.95\columnwidth]{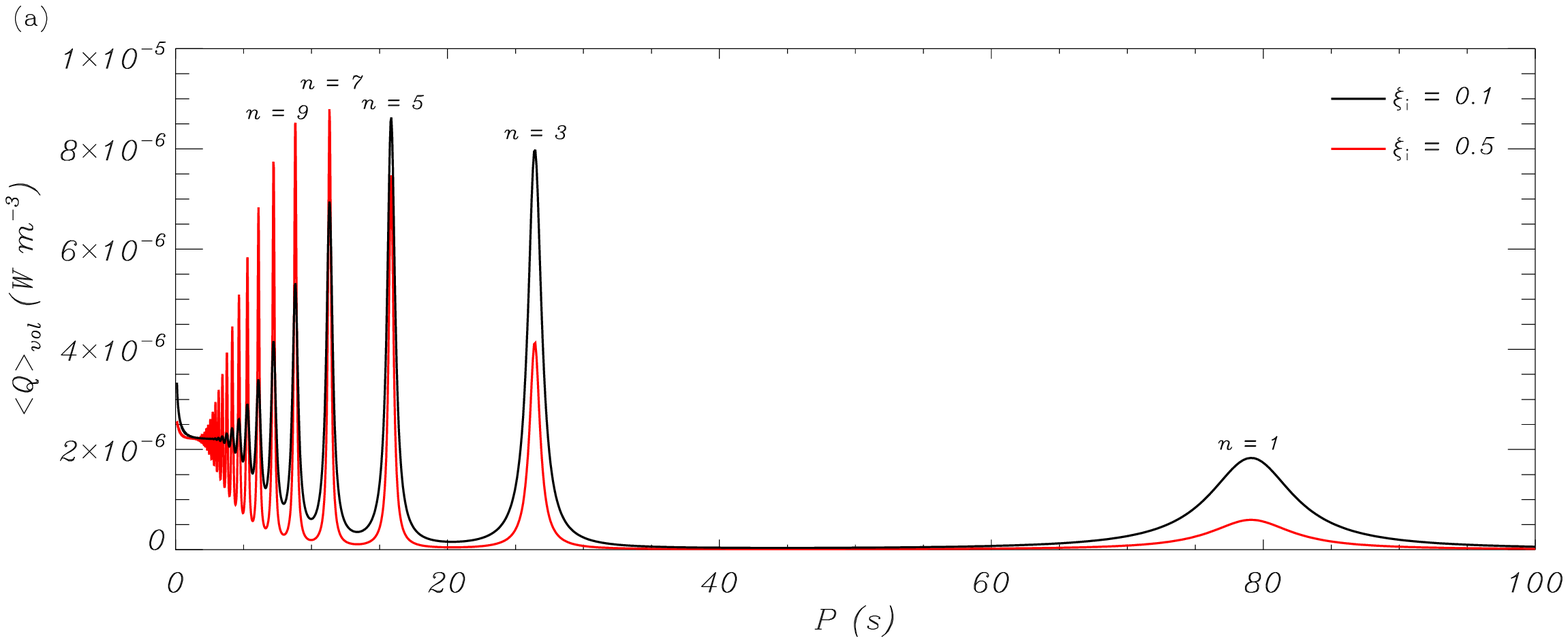}
\includegraphics[width=1.95\columnwidth]{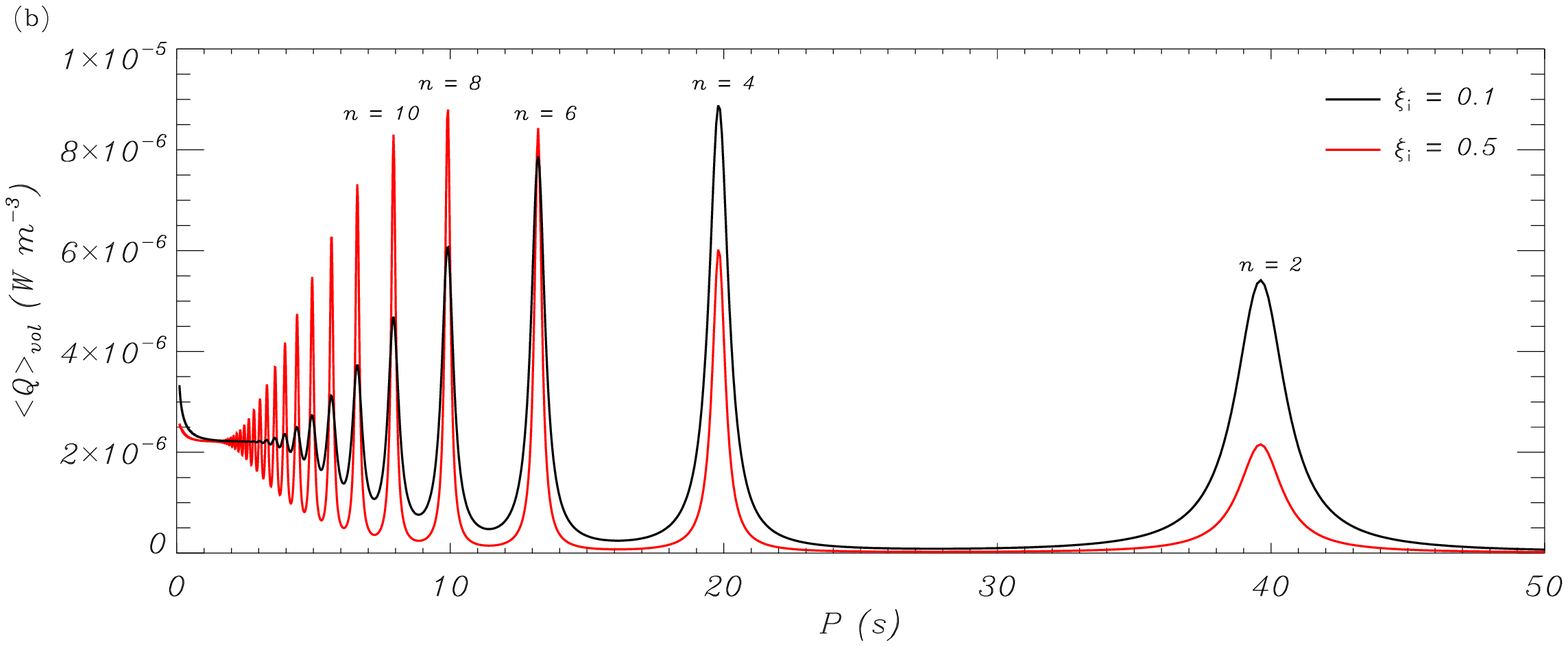}
\caption{Volumetric heating rate,  $\left< Q \right>_{\rm vol}$, as a function of the incident wave period, $P=2\pi/\omega$, for the same physical conditions as in Figure~\ref{fig:slab} with $\xii=0.1$ (black line) and $\xii=0.5$ (red line). In panel (a) we considered $I_1 = -I_2$ so that only odd resonances are present, while in panel (b) we used $I_1 = I_2$ to excite even resonances only. In both cases we consider  10~km~s$^{-1}$ as the velocity amplitude of the incident waves. Note that the horizontal axis range is different in the two panels.
 \label{fig:slab_int}}
\end{figure*}

\begin{figure*}[!htp]
\centering
\includegraphics[width=1.95\columnwidth]{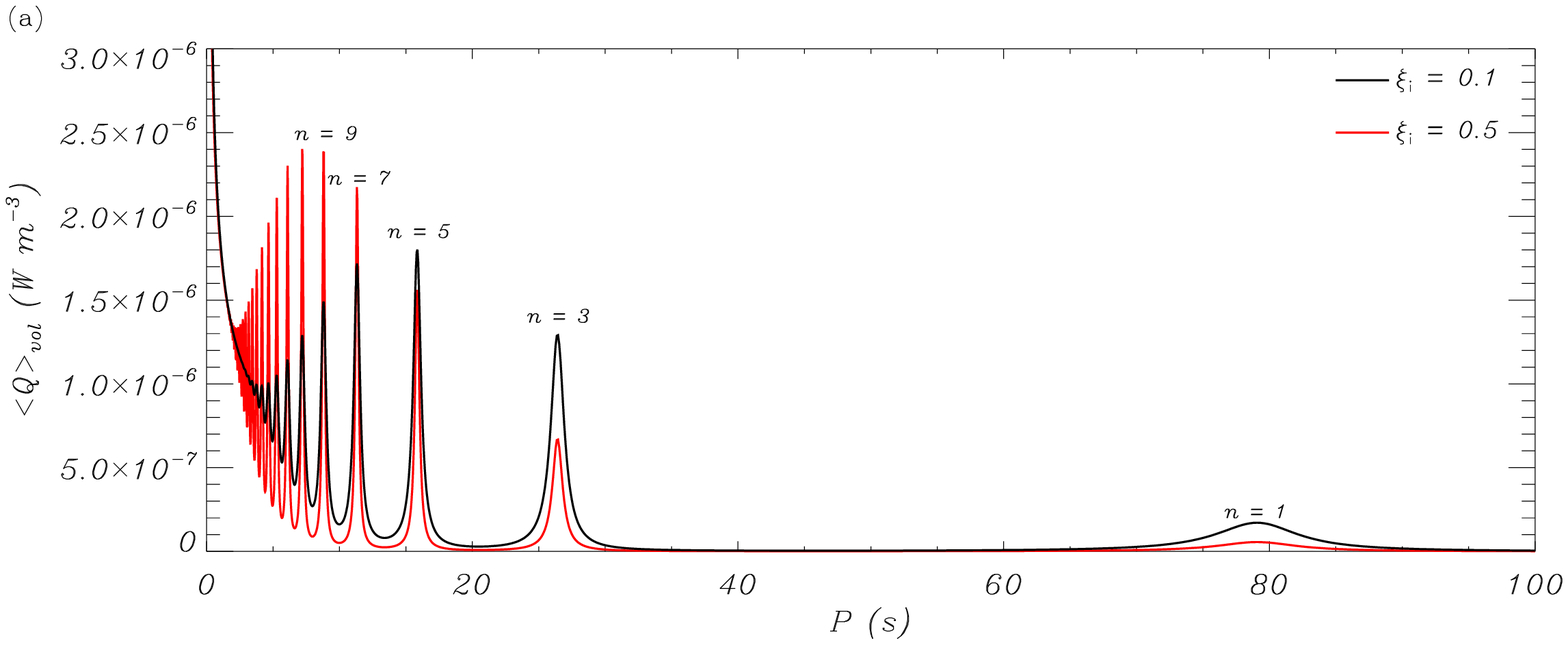}
\includegraphics[width=1.95\columnwidth]{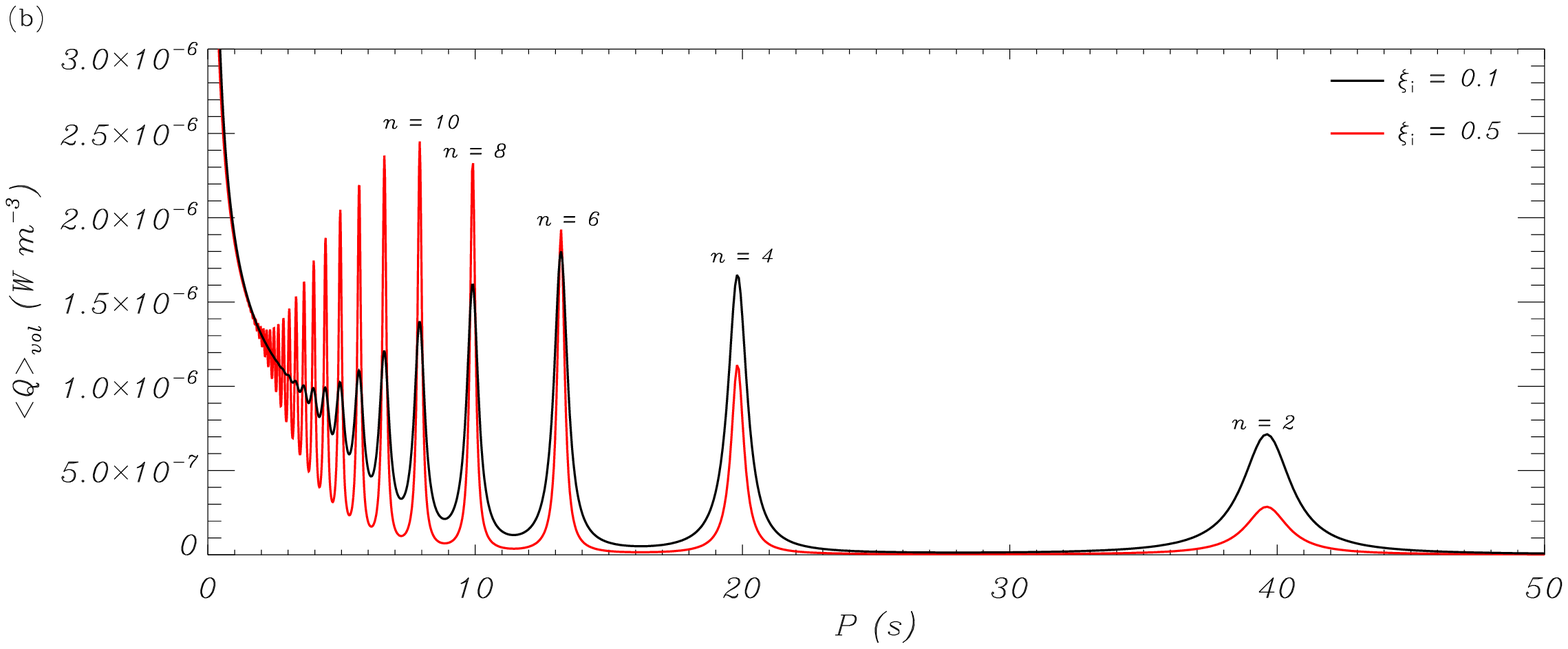}
\caption{Same as Figure~\ref{fig:slab_int} but when the velocity amplitude of the incident waves follows the power law of \citet{hillier2013}.
 \label{fig:slab_int2}}
\end{figure*}

The heating rate caused by the Alfv\'en wave dissipation, $Q$, is given by Equation~(\ref{eq:heat}). Using the ion velocity amplitude of the waves transmitted into the slab given by Equations~(\ref{eq:t1}) and (\ref{eq:t2}), we compute the heating rate within the slab. In order to retain the net heating, we average $Q$ over one wave period to drop the oscillatory dependence.  After some algebraic manipulations, we obtain the expression of the time-averaged  heating, namely
\begin{equation}
\left< Q \right> = \frac{1}{2}\rho_{\rm p} \Lambda \left\{ T_1 T_1^* \exp\left[-2{\rm Im}\left(k_{\rm p}\right)x\right] + T_2 T_2^* \exp\left[2{\rm Im}\left(k_{\rm p}\right)x\right]\right\}, \label{eq:heatingvol}
\end{equation}
with
\begin{eqnarray}
\Lambda &=& \frac{\xii \nuin}{1+4A_{\rm He}/A_{\rm H}} \left| 1 - \frac{q_1}{p} \right|^2 + \frac{\xii \nuine}{1+4A_{\rm He}/A_{\rm H}} \left| 1 - \frac{q_2}{p} \right|^2 \nonumber \\
&& + \frac{\left(1-\xii\right) \nunn}{1+4A_{\rm He}/A_{\rm H}} \left| \frac{q_1-q_2}{p} \right|^2, \\
p &=& \left( \omega + i\nuh \right)\left( \omega + i\nuhe \right) + \nunn\nunne, \\
q_1 &=& i \nuni \left( \omega + i\nuhe \right) - \nunn \nunie, \\
q_2 &=& i \nunie \left( \omega + i\nuh \right) - \nunne \nuni.
\end{eqnarray}
Equation~(\ref{eq:heatingvol}) evidences that the heating profile within the prominence slab is the superposition of two  exponential functions that are maximum at the slab boundaries. Hence, $\left< Q \right>$ is a function of position. The spatial dependence of the heating rate can be dropped by spatially averaging $\left< Q \right> $  within the slab. The result is
\begin{equation}
\left< Q \right>_{\rm vol} = \frac{1}{d}\int_{-d/2}^{d/2} \left< Q \right> {\rm d}x = \rho_{\rm p} \Lambda  \frac{\sinh\left[{\rm Im}\left(k_{\rm p}\right)d\right]}{{\rm Im}\left(k_{\rm p}\right)d} \frac{T_1 T_1^* + T_2 T_2^* }{2}.
\end{equation}
 The units of $\left< Q \right>_{\rm vol}$ are W~m$^{-3}$, so that $\left< Q \right>_{\rm vol}$ is the time-averaged, spatially-averaged volumetric heating rate.  
 
The volumetric heating, $\left< Q \right>_{\rm vol}$,  depends on the parameters of the model, on the wave frequency/period, and on the  velocity amplitude of the incident waves. In fact, as shown in Section~\ref{sec:abs}, for a given model parameters the intrinsic energy absorption depends only on the wave frequency. However, to compute the actual heating  information about the power spectrum of the waves is also needed. The spectral weighting function relating the amplitude of the  waves with their frequency needs to be known, but the power spectrum of the waves that are incident on prominences is not known. This poses an important limitation to estimate the actual heating rate, since the only alternative left is assuming {\em ad hoc} the amplitude of the waves.

 A first possibility is assuming a flat power spectrum and using a constant velocity amplitude. This is done in Figure~\ref{fig:slab_int}, where we represent the volumetric heating as a function of the period when a constant velocity amplitude of 10~km~s$^{-1}$ is used for the incident waves. We use this velocity because it corresponds to the non-thermal velocities for $T\sim 10^4$~K measured by \citet{parenti2007}.  Cavity resonances are clearly seen in Figure~\ref{fig:slab_int} as specific periods for which wave heating is enhanced. We consider two different values of $\xii$ to explore the effect of ion-neutral dissipation. Reducing $\xii$ increases the heating efficiency of the cavity resonances with long periods but reduces the efficiency of the resonances with short periods. Conversely, heating in the range of short periods ($P < 1$~s), where resonances are not present, shows little dependence on $\xii$.
 
 Another possibility is adopting the velocity power law obtained by \citet{hillier2013}, namely $V\sim 10^{0.96} P^{-0.25}$, where $P$ is the period in s and $V$ is the velocity amplitude in km~s$^{-1}$. \citet{hillier2013} derived this power law for transverse (kink) waves observed in prominence threads and with $P > 50$~s. In this application, we assume that the power law is also valid for Alfv\'en waves and remains valid when $P<50$~s. Figure~\ref{fig:slab_int2} shows the corresponding heating rates obtained with this choice of the velocity amplitude. The heating rates associated to cavity resonances  in Figure~\ref{fig:slab_int2} are smaller than those of Figure~\ref{fig:slab_int} because the velocity amplitudes in the intermediate range of periods are smaller than 10~km~s$^{-1}$. For instance, the velocity is  3.5~km~s$^{-1}$ for $P=50$~s and  5.1~km~s$^{-1}$ for $P=10$~s. However, the heating rates in the short-period range are larger in Figure~\ref{fig:slab_int} than in  Figure~\ref{fig:slab_int2}.

\begin{figure*}[!htp]
\centering
\includegraphics[width=.95\columnwidth]{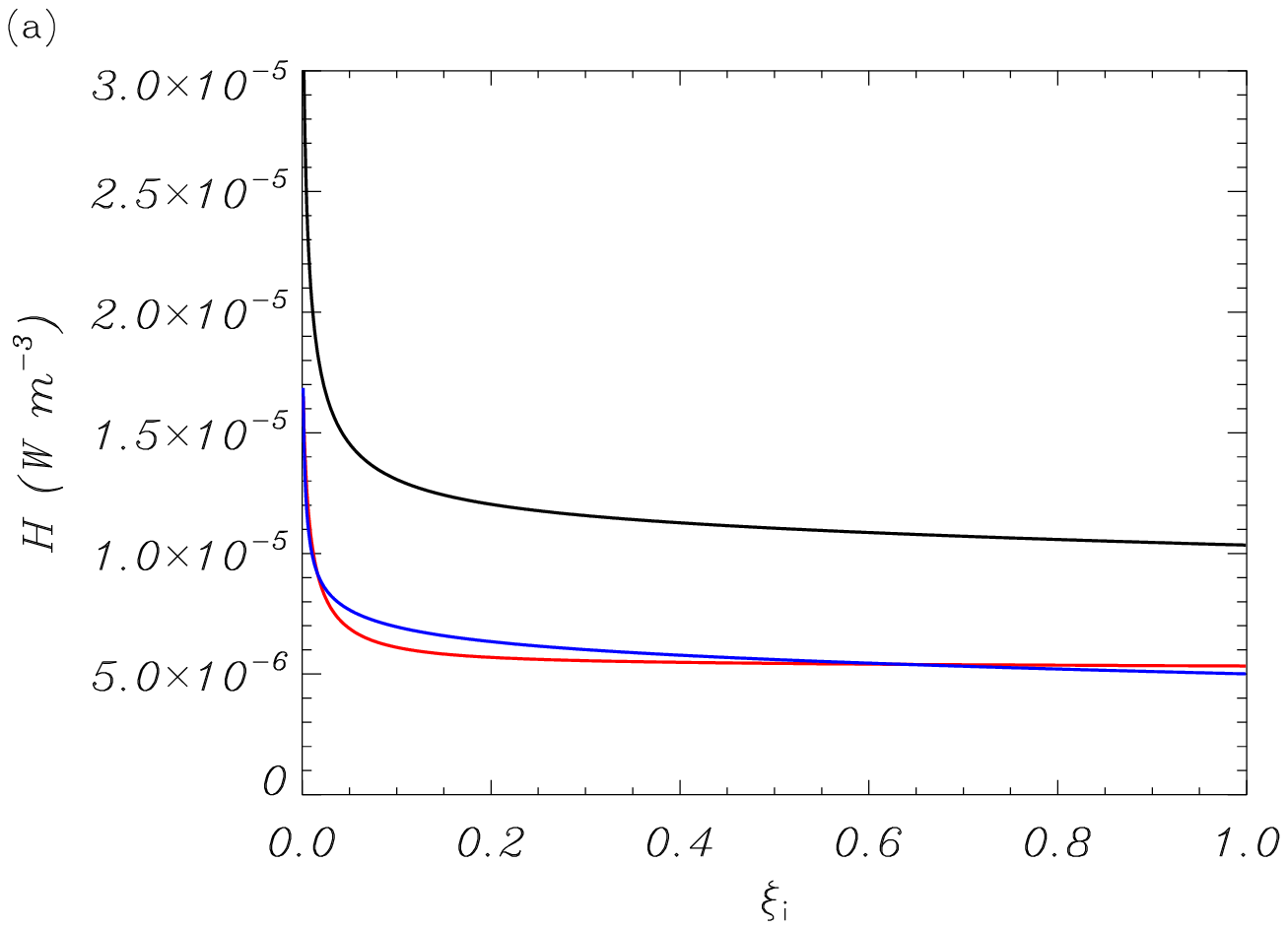}
\includegraphics[width=.95\columnwidth]{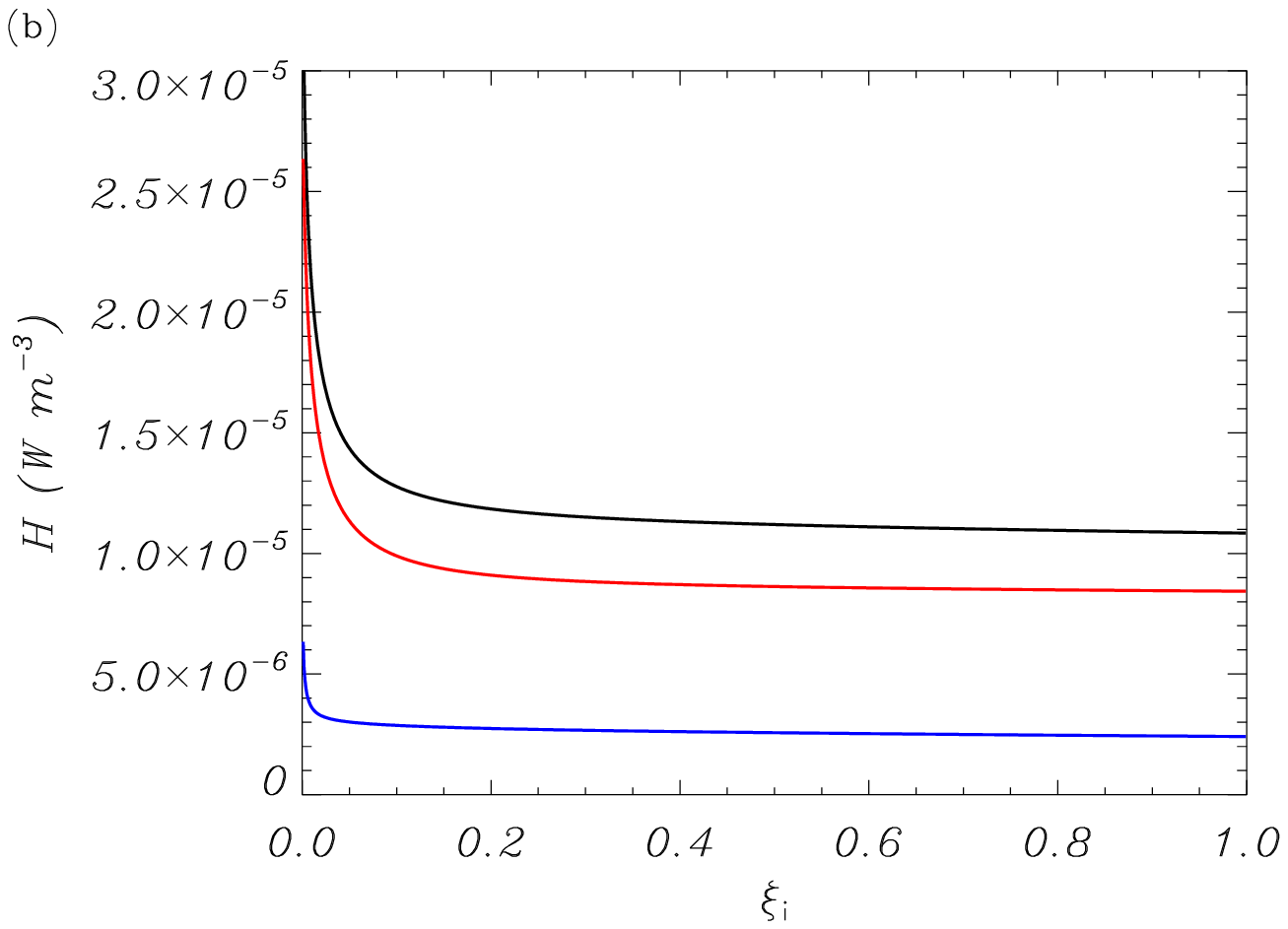}
\caption{(a) Total integrated heating, $H$, as a function of the hydrogen ionization ratio, $\xii$, when the velocity amplitude of the incident waves is 10~km~s$^{-1}$. The black line is the full result, the red line is the contribution of short periods (0.1~s~$\leq P \leq $~1~s), and the blue line is the contribution of intermediate periods (1~s~$\leq P \leq $~100~s)  (b) Same as panel (a) but for the velocity  power law of \citet{hillier2013}.
 \label{fig:heating}}
\end{figure*}

In order to estimate the total heating produced by a broadband driver, we compute the volumetric heating integrated over the range of periods between $P_1$ and $P_2$ as
\begin{equation}
H = \int_{P_1}^{P_2} \frac{1}{P}   \left< Q \right>_{\rm vol} {\rm d} P,
\end{equation}
where we consider $P_1 = 0.1$~s and $P_2=100$~s. This is displayed in Figure~\ref{fig:heating} as a function of  $\xii$, where we considered the two cases of Figures~\ref{fig:slab_int} and \ref{fig:slab_int2}. Importantly, we find that the value of $H$ is independent of the phase relation between $I_1$ and $I_2$. Figure~\ref{fig:heating} shows that the total heating is weakly dependent on the value of  $\xii$ except when the plasma is very weakly ionized ( $\xii \to 0$). In addition, the total heating turns out to be very similar for  the two considered cases, namely $H\sim 10^{-5}$~W~m$^{-3}$, but the  contributions from intermediate (1~s~$\leq P \leq $~100~s) and short (0.1~s~$\leq P \leq $~1~s) periods are different in each case. When the velocity amplitude is 10~km~s$^{-1}$, both intermediate and short periods equally contribute to the total heating. However, when the velocity power law of \citet{hillier2013} is used instead, the contribution from short periods is more important than that from intermediate periods. This is somewhat expected, because  the velocity amplitude increases as the period decreases according to the  power law of \citet{hillier2013}. We also stress that, in both cases, the contribution of intermediate periods mainly comes from cavity resonances, so that the heating associated to those intermediate periods would be much less important if cavity resonances were absent.
 
To assess the relative importance of wave heating, the  heating rate is to be compared to the amount of energy lost by radiation. Determining the  radiative losses of the cool prominence plasma as function of temperature and density is a difficult task that requires complicated numerical solutions of the radiative transfer problem  \citep[e.g.,][]{Anzer1999}.  An easier alternative approach to estimate the radiative losses is based on a semi-empirical parametrization of the radiative loss function \citep[e.g.,][]{hildner}, namely
\begin{equation}
L = \rho_{\rm p}^2\, \chi^* T_{\rm p}^\alpha, \label{eq:losses}
\end{equation}
where $\chi^*$ and $\alpha$ are piecewise functions of the plasma temperature.  An inconvenience of this approach is that Equation~(\ref{eq:losses}) is obtained under the assumption of optically thin plasma, while the cool prominence plasma does not completely satisfy this condition. The uncertainty associated to  Equation~(\ref{eq:losses}) may be large. Different parametrizations of $\chi^*$ and $\alpha$ are available in the literature, which give results that can vary up to an order of magnitude  \citep[see][]{solerparenti2012}. For instance, using the classic fit by \citet{hildner} we compute $L\sim 3\times 10^{-5}$~W~m$^{-3}$ for $\rho_{\rm p} = 5\times 10^{-11}$~kg~m$^{-3}$ and $T_{\rm p}=8000$~K. However,   using the more recent fit based on up-to-date radiative losses computed from the  CHIANTI atomic database by \citet{parenti2006}, and given in \citet{solerparenti2012}, we get $L\sim 1.4\times 10^{-4}$~W~m$^{-3}$ for the same parameters. We consider this last estimation to be more reliable. Another possibility to estimate radiative losses is using the integrated line-of-sight losses computed by \citet{parenti2007} from the differential emission measure. \citet{parenti2007} found that the integrated radiative losses over the temperature range of $10^4$~K--$10^6$~K in a quiescent prominence yielded $3\times 10^3$~W~m$^{-2}$, with most of the emission coming from the low-temperature plasma. To estimate the volumetric losses, we divide that quantity by the width of the prominence slab, $d=$~5,000~km. This results in $6\times 10^{-4}$~W~m$^{-3}$, about 4 times larger than the estimation using Equation~(\ref{eq:losses}) with the CHIANTI-based fit, but it also includes the emission from the plasma at higher temperatures. Finally, another option is to use the radiative losses obtained in actual radiative transfer computations. For instance, Figure~4 of \citet{Heinzel2010} shows radiative losses of the order of $10^{-5}$~W~m$^{-3}$ when losses by hydrogen and calcium are taken into account in a prominence slab with a geometrical width of 5,000~km and a central temperature of 6,800~K. This is an order of magnitude smaller than the result of Equation~(\ref{eq:losses}), although the temperature is also smaller, so that the actual losses for 8,000~K would be presumably larger. Considering the results from these three independent determinations, we can roughly estimate volumetric radiative losses of the  prominence slab to be in between $10^{-5}$~W~m$^{-3}$ and $10^{-4}$~W~m$^{-3}$.

The comparison of the integrated heating rates of Figure~\ref{fig:heating} with the estimated radiative losses indicates that wave heating  may  compensate for a non-negligible fraction of the energy lost by radiation. Considering the large number of parameters involved, among which the power spectrum of the incident waves is particularly important, we prefer to be cautious in our conclusions. An educated guess based on the present results is that wave heating can account for about 10\% of radiative losses. This estimation is in line with the idea that wave heating, if important, should represent a relatively small contribution to the required total heating in prominences \citep{gilbert2015}.

\section{CONCLUSIONS}
\label{sec:conclusion}

In this paper we have explored the role of Alfv\'en waves in prominence heating. Although it is generally accepted that incident radiation basically provides most of the heating to the prominence plasma \citep[see][]{gilbert2015}, results from radiative-equilibrium models indicate that an additional, non-negligible source of heating may be necessary to reproduce the observed temperatures in the prominence cores \citep[e.g.,][]{Heinzel2010,Heinzel2012}. It has been suggested that dissipation of MHD energy may be one of the possible mechanisms \citep{labrosse2010,Heinzel2010,gilbert2015}. Wave heating is a long-hypothesized mechanism for heating of the solar atmospheric plasma, including prominences \citep{arregui2015}.

We have used a slab model with a transverse magnetic field to represent a solar prominence embedded in the corona. We have  modelled the prominence medium as a three-fluid plasma composed of a charged ion-electron single fluid and two neutral fluids made of neutral hydrogen and neutral helium. These three fluids exchange momentum because of particle collisions, which cause the damping of the waves and the dissipation of wave energy.  We have assumed that the driver of the Alfv\'en  waves is located outside the prominence. We have discussed the trapping of energy of Alfv\'en waves incident on the prominence slab and the associated heating rate.

We found that wave energy absorption and associated heating are negligible when the wave period is $P \gtrsim 100$~s. This result has direct observational consequences because the  majority of MHD waves observed in prominences have periods of the order of few minutes \citep[see][]{arreguireview}. Caution is needed when interpreting the role of the observed waves in prominence heating. The evidence that waves are observed propagating in the plasma does not necessarily imply that those waves are actually depositing their energy in the form of heat. Our results indicate that it is unlikely that waves with $P \gtrsim 100$~s have a relevant influence on  prominence energetics. 

Conversely, owing to the existence of cavity resonances \citep[see][]{hollweg1984a}, the trapping of wave energy within the prominence  slab can be very efficient when the period of the incident Alfv\'en waves matches a resonance period.  The channelled energy into the prominence is then dissipated by ion-neutral collisions. For typical prominence parameters, the periods of the cavity resonances appear in the range between 1~s and 100~s. We note that the detection in prominences of periods shorter than 50~s are scarce  \citep[see, e.g.,][]{Balthasar1993,Kolobov2008}, probably because of observational constraints. However, non-thermal velocities derived from line widths have been interpreted as a signature of unresolved waves \citep{parenti2007} and, as \citet{pecseli2000} suggest, short periods can be produced by several mechanisms such as turbulent cascade and/or plasma instabilities.

 We have consistently computed the  plasma heating rate  and have compared it with the prominence radiative losses. There is much uncertainty in this comparison. On  one hand, the power spectrum of the waves that are incident on prominences and their corresponding spectral weighting function are not known, so we were forced to assume the amplitude of the waves when computing the heating rate. On the other hand, we have used a proxy to estimate the radiative losses and  the uncertainty associated to this estimation can be large. Taking into account these limitations, we have concluded that wave heating  may be efficient enough to compensate a fraction of the radiated energy. We  estimate the  volumetric  heating integrated over the range of periods between 0.1~s and 100~s  to be as large as 10\% of the bulk radiative energy of the cool prominence plasma. As expected, the wave heating rate estimated here represents a small contribution to the  total heating necessary to balance emitted radiation, but it can possibly account for the additional heating necessary to explain the observed prominence core temperatures \citep{Heinzel2010}. 
 
 This paper has been a first attempt to understand the actual role of Alfv\'en waves in prominence heating. The main conclusion is that wave heating can play a small but maybe necessary role in prominence energy balance. The obtained results are promising and open the door for more elaborated investigations. This study is restricted by the limitations of linear theory and by the simplicity of the model, so there is room for improvement in both directions.  Future works should investigate the role of wave heating in more realistic 2D or 3D numerical models of prominences, like those developed by, e.g., \citet{terradas2013,terradas2015,terradas2016,Keppens2014,xia2014} among others. In addition, dissipation due to ion-neutral collisions should be included in the numerical models \citep{Hillier2010,Khomenko2014rti,terradasletter2015} as well as the presence of the wave driver at the photosphere. Advanced models should also include the prominence fine structure, i.e., the threads. It is likely that in a threaded prominence the different threads could act as individual Alfv\'enic cavities with their own resonance periods. This may naturally cause the efficiency of wave heating as a function of the wave period to be rather non-uniform within the prominence, so increasing the complexity of the energy balance problem. Other effects not considered here are the heating caused by Alfv\'en wave turbulence \citep[e.g.,][]{vanball2011} or that associated with resonant absorption of transverse waves and generated Kelvin-Helmholtz vortices in the prominence fine structure \citep[e.g.,][]{antolin2015}. These mechanisms can enhance the overall efficiency of wave heating and should also be considered in  future works.

\begin{acknowledgements}
We acknowledge the support  from MINECO and FEDER Funds through grant AYA2014-54485-P. R.S. acknowledges support from MINECO and UIB through a `Ram\'on y Cajal' grant (RYC-2014-14970). J.T. acknowledges support from UIB and the `Ram\'on y Cajal' program.
\end{acknowledgements}

%-------------------------------------------------------------------
\bibliographystyle{aa.bst} % style aa.bst
\bibliography{refs}

\begin{thebibliography}{74}
\expandafter\ifx\csname natexlab\endcsname\relax\def\natexlab#1{#1}\fi

\bibitem[{{Antolin} {et~al.}(2015){Antolin}, {Okamoto}, {De Pontieu},
  {Uitenbroek}, {Van Doorsselaere}, \& {Yokoyama}}]{antolin2015}
{Antolin}, P., {Okamoto}, T.~J., {De Pontieu}, B., {et~al.} 2015, \apj, 809, 72

\bibitem[{{Anzer}(2009)}]{anzer2009}
{Anzer}, U. 2009, \aap, 497, 521

\bibitem[{{Anzer} \& {Heinzel}(1999)}]{Anzer1999}
{Anzer}, U. \& {Heinzel}, P. 1999, \aap, 349, 974

\bibitem[{{Arber} {et~al.}(2016){Arber}, {Brady}, \& {Shelyag}}]{arber2016}
{Arber}, T.~D., {Brady}, C.~S., \& {Shelyag}, S. 2016, \apj, 817, 94

\bibitem[{{Arregui}(2015)}]{arregui2015}
{Arregui}, I. 2015, Philosophical Transactions of the Royal Society of London
  Series A, 373, 20140261

\bibitem[{{Arregui} {et~al.}(2012){Arregui}, {Oliver}, \&
  {Ballester}}]{arreguireview}
{Arregui}, I., {Oliver}, R., \& {Ballester}, J.~L. 2012, Living Reviews in
  Solar Physics, 9

\bibitem[{{Ballester}(2015)}]{ballester2015}
{Ballester}, J.~L. 2015, in Astrophysics and Space Science Library, Vol. 415,
  Solar Prominences, ed. J.-C. {Vial} \& O.~{Engvold}, 259

\bibitem[{{Balthasar} {et~al.}(1993){Balthasar}, {Wiehr}, {Schleicher}, \&
  {Wohl}}]{Balthasar1993}
{Balthasar}, H., {Wiehr}, E., {Schleicher}, H., \& {Wohl}, H. 1993, \aap, 277,
  635

\bibitem[{{Braginskii}(1965)}]{brag}
{Braginskii}, S.~I. 1965, Reviews of Plasma Physics, 1, 205

\bibitem[{{Carbonell} {et~al.}(2010){Carbonell}, {Forteza}, {Oliver}, \&
  {Ballester}}]{carbonell2010}
{Carbonell}, M., {Forteza}, P., {Oliver}, R., \& {Ballester}, J.~L. 2010, \aap,
  515, A80

\bibitem[{{Carlsson} \& {Leenaarts}(2012)}]{carlsson2012}
{Carlsson}, M. \& {Leenaarts}, J. 2012, \aap, 539, A39

\bibitem[{{De Pontieu} {et~al.}(2001){De Pontieu}, {Martens}, \&
  {Hudson}}]{depontieu2001}
{De Pontieu}, B., {Martens}, P.~C.~H., \& {Hudson}, H.~S. 2001, \apj, 558, 859

\bibitem[{{Draine}(1986)}]{draine86}
{Draine}, B.~T. 1986, \mnras, 220, 133

\bibitem[{{Engvold}(2015)}]{engvold2015}
{Engvold}, O. 2015, in Astrophysics and Space Science Library, Vol. 415, Solar
  Prominences, ed. J.-C. {Vial} \& O.~{Engvold}, 31

\bibitem[{{Ferraro}(1954)}]{ferraro1954}
{Ferraro}, V.~C.~A. 1954, \apj, 119, 393

\bibitem[{{Forteza} {et~al.}(2007){Forteza}, {Oliver}, {Ballester}, \&
  {Khodachenko}}]{forteza2007}
{Forteza}, P., {Oliver}, R., {Ballester}, J.~L., \& {Khodachenko}, M.~L. 2007,
  \aap, 461, 731

\bibitem[{{Gilbert}(2015)}]{gilbert2015}
{Gilbert}, H. 2015, in Astrophysics and Space Science Library, Vol. 415, Solar
  Prominences, ed. J.-C. {Vial} \& O.~{Engvold}, 157

\bibitem[{{Gilbert} {et~al.}(2002){Gilbert}, {Hansteen}, \&
  {Holzer}}]{Gilbert2002}
{Gilbert}, H.~R., {Hansteen}, V.~H., \& {Holzer}, T.~E. 2002, \apj, 577, 464

\bibitem[{{Goedbloed} \& {Poedts}(2004)}]{goedbloed2004}
{Goedbloed}, J.~P. \& {Poedts}, S. 2004, {Principles of magnetohydrodynamics}
  (Cambridge University Press)

\bibitem[{{Goodman}(2011)}]{goodman2011}
{Goodman}, M.~L. 2011, \apj, 735, 45

\bibitem[{{Goossens} {et~al.}(2011){Goossens}, {Erd{\'e}lyi}, \&
  {Ruderman}}]{goossens2011}
{Goossens}, M., {Erd{\'e}lyi}, R., \& {Ruderman}, M.~S. 2011, \ssr, 158, 289

\bibitem[{{Haerendel}(1992)}]{haerendel1992}
{Haerendel}, G. 1992, \nat, 360, 241

\bibitem[{{Heasley} \& {Mihalas}(1976)}]{Heasley1976}
{Heasley}, J.~N. \& {Mihalas}, D. 1976, \apj, 205, 273

\bibitem[{{Heinzel}(2015)}]{heinzel2015}
{Heinzel}, P. 2015, in Astrophysics and Space Science Library, Vol. 415, Solar
  Prominences, ed. J.-C. {Vial} \& O.~{Engvold}, 103

\bibitem[{{Heinzel} \& {Anzer}(2012)}]{Heinzel2012}
{Heinzel}, P. \& {Anzer}, U. 2012, \aap, 539, A49

\bibitem[{{Heinzel} {et~al.}(2010){Heinzel}, {Anzer}, \&
  {Gun{\'a}r}}]{Heinzel2010}
{Heinzel}, P., {Anzer}, U., \& {Gun{\'a}r}, S. 2010, \memsai, 81, 654

\bibitem[{{Hildner}(1974)}]{hildner}
{Hildner}, E. 1974, \solphys, 35, 123

\bibitem[{{Hillier} {et~al.}(2013){Hillier}, {Morton}, \&
  {Erd{\'e}lyi}}]{hillier2013}
{Hillier}, A., {Morton}, R.~J., \& {Erd{\'e}lyi}, R. 2013, \apjl, 779, L16

\bibitem[{{Hillier} {et~al.}(2010){Hillier}, {Shibata}, \&
  {Isobe}}]{Hillier2010}
{Hillier}, A., {Shibata}, K., \& {Isobe}, H. 2010, \pasj, 62, 1231

\bibitem[{{Hollweg}(1981)}]{hollweg1981}
{Hollweg}, J.~V. 1981, \solphys, 70, 25

\bibitem[{{Hollweg}(1984{\natexlab{a}})}]{hollweg1984a}
{Hollweg}, J.~V. 1984{\natexlab{a}}, \solphys, 91, 269

\bibitem[{{Hollweg}(1984{\natexlab{b}})}]{hollweg1984b}
{Hollweg}, J.~V. 1984{\natexlab{b}}, \apj, 277, 392

\bibitem[{{Ionson}(1982)}]{ionson1982}
{Ionson}, J.~A. 1982, \apj, 254, 318

\bibitem[{{Jensen}(1983)}]{jensen1983}
{Jensen}, E. 1983, \solphys, 89, 275

\bibitem[{{Jensen}(1986)}]{jensen1986}
{Jensen}, E. 1986, in NASA Conference Publication, Vol. 2442, NASA Conference
  Publication, ed. A.~I. {Poland}

\bibitem[{{Joarder} \& {Roberts}(1992)}]{joarder1992}
{Joarder}, P.~S. \& {Roberts}, B. 1992, \aap, 261, 625

\bibitem[{{Keppens} \& {Xia}(2014)}]{Keppens2014}
{Keppens}, R. \& {Xia}, C. 2014, \apj, 789, 22

\bibitem[{{Khodachenko} {et~al.}(2004){Khodachenko}, {Arber}, {Rucker}, \&
  {Hanslmeier}}]{khodachenko2004}
{Khodachenko}, M.~L., {Arber}, T.~D., {Rucker}, H.~O., \& {Hanslmeier}, A.
  2004, \aap, 422, 1073

\bibitem[{{Khodachenko} {et~al.}(2006){Khodachenko}, {Rucker}, {Oliver},
  {Arber}, \& {Hanslmeier}}]{khodachenko2006}
{Khodachenko}, M.~L., {Rucker}, H.~O., {Oliver}, R., {Arber}, T.~D., \&
  {Hanslmeier}, A. 2006, Advances in Space Research, 37, 447

\bibitem[{{Khomenko} {et~al.}(2014{\natexlab{a}}){Khomenko}, {Collados},
  {D{\'{\i}}az}, \& {Vitas}}]{Khomenko2014}
{Khomenko}, E., {Collados}, M., {D{\'{\i}}az}, A., \& {Vitas}, N.
  2014{\natexlab{a}}, Physics of Plasmas, 21, 092901

\bibitem[{{Khomenko} {et~al.}(2014{\natexlab{b}}){Khomenko}, {D{\'{\i}}az}, {de
  Vicente}, {Collados}, \& {Luna}}]{Khomenko2014rti}
{Khomenko}, E., {D{\'{\i}}az}, A., {de Vicente}, A., {Collados}, M., \& {Luna},
  M. 2014{\natexlab{b}}, \aap, 565, A45

\bibitem[{{Kolobov} {et~al.}(2008){Kolobov}, {Kobanov}, \&
  {Chupin}}]{Kolobov2008}
{Kolobov}, D.~Y., {Kobanov}, N.~I., \& {Chupin}, S.~A. 2008, in IAU Symposium,
  Vol. 247, Waves \& Oscillations in the Solar Atmosphere: Heating and
  Magneto-Seismology, ed. R.~{Erd{\'e}lyi} \& C.~A. {Mendoza-Briceno}, 178--181

\bibitem[{{Labrosse} {et~al.}(2010){Labrosse}, {Heinzel}, {Vial}, {Kucera},
  {Parenti}, {Gun{\'a}r}, {Schmieder}, \& {Kilper}}]{labrosse2010}
{Labrosse}, N., {Heinzel}, P., {Vial}, J.-C., {et~al.} 2010, \ssr, 151, 243

\bibitem[{{Leake} {et~al.}(2005){Leake}, {Arber}, \& {Khodachenko}}]{leake2005}
{Leake}, J.~E., {Arber}, T.~D., \& {Khodachenko}, M.~L. 2005, \aap, 442, 1091

\bibitem[{{Lewkow} {et~al.}(2012){Lewkow}, {Kharchenko}, \&
  {Zhang}}]{Lewkow2012}
{Lewkow}, N.~R., {Kharchenko}, V., \& {Zhang}, P. 2012, \apj, 756, 57

\bibitem[{{Lin} {et~al.}(2007){Lin}, {Engvold}, {Rouppe van der Voort}, \& {van
  Noort}}]{lin2007}
{Lin}, Y., {Engvold}, O., {Rouppe van der Voort}, L.~H.~M., \& {van Noort}, M.
  2007, \solphys, 246, 65

\bibitem[{{Lin} {et~al.}(2009){Lin}, {Soler}, {Engvold}, {Ballester},
  {Langangen}, {Oliver}, \& {Rouppe van der Voort}}]{lin2009}
{Lin}, Y., {Soler}, R., {Engvold}, O., {et~al.} 2009, \apj, 704, 870

\bibitem[{{Oliver} {et~al.}(1993){Oliver}, {Ballester}, {Hood}, \&
  {Priest}}]{oliver1993}
{Oliver}, R., {Ballester}, J.~L., {Hood}, A.~W., \& {Priest}, E.~R. 1993, \apj,
  409, 809

\bibitem[{{Parenti}(2014)}]{parenti2014}
{Parenti}, S. 2014, Living Reviews in Solar Physics, 11

\bibitem[{{Parenti} {et~al.}(2006){Parenti}, {Buchlin}, {Cargill}, {Galtier},
  \& {Vial}}]{parenti2006}
{Parenti}, S., {Buchlin}, E., {Cargill}, P.~J., {Galtier}, S., \& {Vial}, J.-C.
  2006, \apj, 651, 1219

\bibitem[{{Parenti} \& {Vial}(2007)}]{parenti2007}
{Parenti}, S. \& {Vial}, J.-C. 2007, \aap, 469, 1109

\bibitem[{{P{\'e}cseli} \& {Engvold}(2000)}]{pecseli2000}
{P{\'e}cseli}, H. \& {Engvold}, O. 2000, \solphys, 194, 73

\bibitem[{{Reep} \& {Russell}(2016)}]{reep2016}
{Reep}, J.~W. \& {Russell}, A.~J.~B. 2016, \apjl, 818, L20

\bibitem[{{Russell} \& {Fletcher}(2013)}]{russell2013}
{Russell}, A.~J.~B. \& {Fletcher}, L. 2013, \apj, 765, 81

\bibitem[{{Soler} {et~al.}(2012){Soler}, {Ballester}, \&
  {Parenti}}]{solerparenti2012}
{Soler}, R., {Ballester}, J.~L., \& {Parenti}, S. 2012, \aap, 540, A7

\bibitem[{{Soler} {et~al.}(2015{\natexlab{a}}){Soler}, {Ballester}, \&
  {Zaqarashvili}}]{soler2015}
{Soler}, R., {Ballester}, J.~L., \& {Zaqarashvili}, T.~V. 2015{\natexlab{a}},
  \aap, 573, A79

\bibitem[{{Soler} {et~al.}(2015{\natexlab{b}}){Soler}, {Carbonell}, \&
  {Ballester}}]{soler2015apj}
{Soler}, R., {Carbonell}, M., \& {Ballester}, J.~L. 2015{\natexlab{b}}, \apj,
  810, 146

\bibitem[{{Soler} {et~al.}(2009{\natexlab{a}}){Soler}, {Oliver}, \&
  {Ballester}}]{soler2009NewA}
{Soler}, R., {Oliver}, R., \& {Ballester}, J.~L. 2009{\natexlab{a}}, \na, 14,
  238

\bibitem[{{Soler} {et~al.}(2009{\natexlab{b}}){Soler}, {Oliver}, \&
  {Ballester}}]{soler2009PI}
{Soler}, R., {Oliver}, R., \& {Ballester}, J.~L. 2009{\natexlab{b}}, \apj, 699,
  1553

\bibitem[{{Soler} {et~al.}(2009{\natexlab{c}}){Soler}, {Oliver}, \&
  {Ballester}}]{soler2009PIRA}
{Soler}, R., {Oliver}, R., \& {Ballester}, J.~L. 2009{\natexlab{c}}, \apj, 707,
  662

\bibitem[{{Song} \& {Vasyli{\= u}nas}(2011)}]{song2011}
{Song}, P. \& {Vasyli{\= u}nas}, V.~M. 2011, Journal of Geophysical Research
  (Space Physics), 116, 9104

\bibitem[{{Sterling} \& {Hollweg}(1984)}]{Sterling1984}
{Sterling}, A.~C. \& {Hollweg}, J.~V. 1984, \apj, 285, 843

\bibitem[{{Terradas} {et~al.}(2013){Terradas}, {Soler}, {D{\'{\i}}az},
  {Oliver}, \& {Ballester}}]{terradas2013}
{Terradas}, J., {Soler}, R., {D{\'{\i}}az}, A.~J., {Oliver}, R., \&
  {Ballester}, J.~L. 2013, \apj, 778, 49

\bibitem[{{Terradas} {et~al.}(2015{\natexlab{a}}){Terradas}, {Soler}, {Luna},
  {Oliver}, \& {Ballester}}]{terradas2015}
{Terradas}, J., {Soler}, R., {Luna}, M., {Oliver}, R., \& {Ballester}, J.~L.
  2015{\natexlab{a}}, \apj, 799, 94

\bibitem[{{Terradas} {et~al.}(2016){Terradas}, {Soler}, {Luna}, {Oliver},
  {Ballester}, \& {Wright}}]{terradas2016}
{Terradas}, J., {Soler}, R., {Luna}, M., {et~al.} 2016, \apj, 820, 125

\bibitem[{{Terradas} {et~al.}(2015{\natexlab{b}}){Terradas}, {Soler}, {Oliver},
  \& {Ballester}}]{terradasletter2015}
{Terradas}, J., {Soler}, R., {Oliver}, R., \& {Ballester}, J.~L.
  2015{\natexlab{b}}, \apjl, 802, L28

\bibitem[{{Tu} \& {Song}(2013)}]{tu2013}
{Tu}, J. \& {Song}, P. 2013, \apj, 777, 53

\bibitem[{{van Ballegooijen} {et~al.}(2011){van Ballegooijen}, {Asgari-Targhi},
  {Cranmer}, \& {DeLuca}}]{vanball2011}
{van Ballegooijen}, A.~A., {Asgari-Targhi}, M., {Cranmer}, S.~R., \& {DeLuca},
  E.~E. 2011, \apj, 736, 3

\bibitem[{{Vranjes} \& {Krstic}(2013)}]{vranjes2013}
{Vranjes}, J. \& {Krstic}, P.~S. 2013, \aap, 554, A22

\bibitem[{{Walker}(2005)}]{walker2005}
{Walker}, A.~D.~M. 2005, {Magnetohydrodynamic Waves in Geospace}, Series in
  Plasma Physics, (Institute of Physics Publishing)

\bibitem[{{Xia} {et~al.}(2014){Xia}, {Keppens}, \& {Guo}}]{xia2014}
{Xia}, C., {Keppens}, R., \& {Guo}, Y. 2014, \apj, 780, 130

\bibitem[{{Zaqarashvili} {et~al.}(2011{\natexlab{a}}){Zaqarashvili},
  {Khodachenko}, \& {Rucker}}]{Zaqarashvili2011b}
{Zaqarashvili}, T.~V., {Khodachenko}, M.~L., \& {Rucker}, H.~O.
  2011{\natexlab{a}}, \aap, 534, A93

\bibitem[{{Zaqarashvili} {et~al.}(2011{\natexlab{b}}){Zaqarashvili},
  {Khodachenko}, \& {Rucker}}]{zaqarashvili2011a}
{Zaqarashvili}, T.~V., {Khodachenko}, M.~L., \& {Rucker}, H.~O.
  2011{\natexlab{b}}, \aap, 529, A82

\bibitem[{{Zhugzhda} \& {Locans}(1982)}]{Zhugzhda1982}
{Zhugzhda}, I.~D. \& {Locans}, V. 1982, \solphys, 76, 77

\end{thebibliography}

\end{document}